\documentstyle[emulapj,epsfig]{article}

\font\tenbg=cmmib10 at 10pt
\def \rvecmu{{\hbox{\tenbg\char'026}}}

\begin{document}

\title{
``Propeller" Regime of Disk Accretion to  Rapidly Rotating Stars }

\author{G.V.~Ustyugova}
\affil{Keldysh Institute of Applied Mathematics, Russian Academy of
Sciences, Moscow, Russia;~ ustyugg@spp.Keldysh.ru}

\author{A.V.~Koldoba}
\affil{Institute of Mathematical Modeling, Russian Academy of
Sciences, Moscow, Russia;~koldoba@spp.Keldysh.ru}

\author{M.M.~Romanova}
\affil{Department of Astronomy, Cornell University, Ithaca, NY
14853-6801; ~ romanova@astro.cornell.edu}

\author{R.V.E.~Lovelace}
\affil{Department of Astronomy, Department of Applied and
Engineering Physics, Cornell University, Ithaca, NY 14853-6801;
~RVL1@cornell.edu}


\begin{abstract}

       We present  results of
axisymmetric
 magnetohydrodynamic simulations of the interaction of a
rapidly rotating,
magnetized star with an accretion disk.
          The disk is considered to have a
finite viscosity and magnetic diffusivity.
         The main parameters of the system are
the star's angular velocity and magnetic moment, and the disk's
viscosity
and
  diffusivity.
        We focus on the ``propeller" regime
where the inner radius of the disk is larger than the corotation
radius.
         Two types
of magnetohydrodynamic flows have been found as a result of
simulations: ``weak" and ``strong" propellers.
         The strong
propellers are
characterized by a powerful disk wind and a collimated magnetically
dominated outflow or jet from the star. The weak
propellers
have only weak outflows. We investigated the time-averaged
characteristics of the interaction between the main elements of the
system,  the star, the disk, the wind from the disk, and the jet.
        Rates of exchange of mass and angular momentum
between the elements of the system are derived as a function of the
main parameters. The propeller mechanism may be responsible for the
fast spinning-down of
the
classical T Tauri stars in the initial stages of their evolution,
and for the spinning-down of accreting millisecond pulsars.

\end{abstract}

\section{Introduction}

This work studies the interaction of a rapidly rotating magnetized
star with an accretion disk under conditions where the corotation
radius of the star, ${r_{c}=( {GM}/{\Omega_*^2} )^{1/3}}$, is less
than the inner radius of the disk, $r_{d}$.
         The radius $r_d$ is determined by the star's
magnetic field and angular velocity as well as the  mass accretion
rate of the disk
       $\dot{M}$ and the disk viscosity and magnetic diffusivity.
          Disk accretion is disrupted for radii $\lesssim r_d$.
        At distances $\sim r_d$ the disk plasma
acquires additional angular velocity due to ``friction" with the
magnetosphere of the star which rotates faster than the Keplerian
motion of the disk.
          The behavior of the matter at $r_d$
depends  on the ratio between angular velocity of
       the star and Keplerian angular velocity at $r_d$.
          The matter  tends to be expelled from
the disk if the centrifugal force is sufficiently larger than the
gravitational force.
        This regime of interaction between
a magnetized star and an accretion disk is called ``propeller"
regime (Illarionov \& Sunyaev 1975).
This regime was investigated analytically  (e.g., Davies, Fabian, \&
Pringle 1979; Lovelace, Romanova, \& Bisnovatyi-Kogan 1999,
hereafter - LRBK99; Ikhsanov 2002;
 Rappaport, Fregeau,
\& Spruit 2004; Eksi, Hernquist, \& Narayan 2005), and numerically (
Romanova et al. 2004b - hereafter RUKL04; Romanova et al. 2005 -
hereafter RUKL05).


In  the disk-magnetosphere interaction the magnetic field plays the
main role in two processes.
        Firstly, it disrupts the inner regions of the disk and
thus determines the inner radius of the disk.
        Secondly, the magnetic field  determines
the direction of plasma flow.
         Matter of the  disk at
a distance $r$ is accelerated in the azimuthal direction if it is
threaded by field lines connecting it with a star rotating more
rapidly than the disk; that is $r> r_c$.
        In the  opposite case
where $r<r_c$, the azimuthal motion of the disk matter is slowed
down.
         In the first case the  disk plasma
gains angular momentum from the star and moves outward.
       In second
case the disk plasma loses angular momentum and moves inward. Thus,
in the inner regions of the disk the action of magnetic field  leads
to enhanced accretion rate, while in the outer regions to reduced
accretion
       or even to outflow of matter.

         The disk-magnetosphere interaction
depends essentially on the ratio of the inner radius of the disk
$r_d$ to the corotation radius $r_c$.
         In the model studied here
the location of the radius $r_d$ depends mainly on the magnetic
field of the star and the disk accretion rate.
        In the absence of external plasma
the magnetic field of the star is considered to be a dipole field
with the magnetic moment $\rvecmu$  parallel to the rotational axis.
        Consequently the magnetic stresses
are proportional to $r^{-6}$.

         For $r_d
\lesssim r_c$, disruption of the disk is accompanied by the
formation of ``funnel flows''
(e.g., Ghosh \& Lamb 1978; K\"onigl 1991; Shu et al. 1994).
Funnel flows were recently investigated numerically using 2D and 3D
MHD simulations (Romanova et al. 2002, 2003, 2004a).
         In this paper we consider the opposite
limit where $r_d\gtrsim r_c$.
        In this limit one  expects that
the closed field region of the star will contain relatively
low-density plasma rotating with the star's angular velocity.
         Outside of this region there
is
 a relatively dense disk plasma  threaded by open magnetic field
lines.
          Interaction between
the two regions occurs by two processes: (1) by matter flow from
open field lines to the closed field lines  as a result of magnetic
diffusivity; and (2) angular momentum exchange between fast rotating
matter of magnetosphere and slowly rotating disk.
           The first process can lead to penetration of significant
amounts of matter with low angular momentum to the closed field
lines.
         This can in turn lead to
significant deformation of the field lines and
possibly to their opening.
          The second process
may lead  to azimuthal acceleration of the disk matter sufficient to
stop the accretion and to eject  matter from the disk. An outflow
from the system may occur.

         This paper treats  disk accretion to
a rotating magnetized star using magnetohydrodynamics.
          For the initial set up of the system
we divide the space into the ``disk'' and the ``corona''
       with a smooth transition between them. The disk is
       relatively cold and dense while the opposite is true for
the corona. The disk and corona are separated by
a specified
level of density. In the disk we take into account viscosity and
magnetic diffusivity of the plasma.

        We consider that both the viscosity
and the magnetic diffusivity of the disk plasma are due to
turbulent fluctuations of the velocity and magnetic field.
         We adopt the standard hypotheses
where the  molecular transport coefficients are  substituted with
the turbulent coefficients.
        To estimate the value of these coefficients, we
use the  $\alpha$-model of Shakura and Sunyaev (1973) where the
coefficient of the turbulent kinematic viscosity $\nu_{\rm t} =
\alpha_{\rm v} c_s^2/\Omega_K$, where $c_s$ is the isothermal sound
speed and $\Omega_K$ is the Keplerian angular velocity at the given
location.
         Similarly, the coefficient of the turbulent magnetic
diffusivity $\eta_{\rm t}=\alpha_{\rm d} c_s^2/\Omega_K$. Here,
       $\alpha_{\rm v}$ and $\alpha_{\rm d}$ are dimensionless coefficients which
are treated as parameters of the model.

Our numerical simulations have shown that the MHD flows which appear
as a result of interaction between a rapidly rotating magnetized
star and accretion disk can be divided to two types: ``weak'' and
``strong'' propeller. In the first case, no outflows are observed
(RUKL04),
while in the second case
a large
fraction of the mass accretion of the disk goes to the outflows
(RUKL05). In this paper we investigate regime of ``strong"
propellers in detail.

        The inner disk radius $r_d$ is
determined by the disk interaction with the magnetized star.
         The value $r_d$ varies with
time significantly.
          We can consider however an
average value of $r_d$.
         This value  depends
on the parameters of the model which characterize both the star, its
mass $M$, magnetic moment $\mu$, and angular velocity $\Omega_*$;
and the disk, the coefficients of viscosity and magnetic
diffusivity, density, temperature and connected with them the
accretion rate in the disk.
        The corona may also have a significant role.
            In order to decrease the influence
of corona we have taken its density to be as low as possible.

         From the parameters of
the model we can construct in addition to the corotation radius
another important length termed the nominal Alfv\'en radius,
$r_A\equiv [ {\mu^4}/({GM \dot{M}^2})]^{1/7}$
(e.g., Davidson \& Ostriker 1973).
 Here, $\dot{M}$ is the
disk accretion rate at large distances, which can be obtained from
the simulations.
         Another characteristic length  can
be obtained from the following.
         Close to the star
the magnetic field dominates, that is, the magnetic pressure is
larger than the $\phi \phi$ component of the momentum tensor of the
matter $\rho v_\phi^2+p$.
        Inside the
region where the magnetic field dominates, the plasma flow is
controlled by the magnetic field and the flow is  along the magnetic
field lines.
The
boundary of this region follows from $p+\rho v_\phi^2= {\bf B}^2/8
\pi$.
         This surface intersects the equatorial region
along the circle with the radius $r_m$, which we will call the
magnetospheric radius.

        The strength of the propeller is determined by the
value of the dimensionless ratio, $r_A/r_c$ or $r_m/r_c$, and also
by other dimensionless parameters of the model such
as
 the $\alpha$-coefficients of the viscosity and magnetic
diffusivity.
        The characteristic feature of the
disk-magnetosphere interaction in the strong propeller regime is
the formation of an outflow   of the disk matter along the opened
magnetic field lines.
        This outflow is termed the ``wind.''
         The matter outflows are strongest
in the vicinity of the neutral line of the poloidal field, which
starts approximately at the inner radius of the disk.
         This is the region of the most intense angular momentum
transport from the fast rotating magnetosphere to the disk due to
``viscous" stresses.
          At the same time there is a magnetically dominated
(Poynting) ``jet''  which flows out along the opened magnetic field
lines extending outward from the star.
         The wind has a large inclination angle relative to the rotational
axis and for a very strong propeller it flows almost along the disk
surface.
           In contrast, the jet becomes more  collimated as
the strength of the propeller increases.
         As the strength increases there is
a sharp decrease of the accretion rate to the star.

Section 2 discusses theoretical approach to the problem. Section 3
discusses our numerical model.
Section 4 discusses results and Section 5 summarizes the work.

\section{Structure and Evolution of the Disk with Magnetic Field}

\subsection{Non-Magnetic Disk}

        For a stationary non-magnetic thin disk, the fluxes
of matter  $\dot{M}$ and angular momentum $\dot{L}$ are conserved
along the disk (see, e.g.,  Kato, Fukue \& Mineshige 1988; Lipunov
1992)
\begin{equation}\label{eq1}
       \begin{array}{ccccll}
         r \Sigma V_r & = & -\displaystyle{\frac{\dot{M}}{2\pi} } = {\rm
const} ~,
         \\[0.3cm]
         r^2 \left( \Sigma V_r V_\phi +\tau \right) & = &
         \displaystyle{-\frac{\dot{L}}{2\pi}} =  {\rm const}  ~.\\
       \end{array}
\end{equation}
Here, $\Sigma = \int  dz~\rho $ is the surface density; $\tau = \int
dz~\tau_{r \phi} $ is the ``viscous'' stress acting in the disk,
$V_r$ and $V_\phi$ are components of the flow velocity.
       Here, $\dot{M}>0$ is the inward mass flux, and $\dot L>0$
is the inward angular momentum flux.
       From equations (\ref{eq1}) we obtain
\begin{equation}\label{eq2}
       \tau= \frac{1}{2\pi} \left( \dot M \Omega - \frac{\dot L}{r^2}
       \right) =
       \frac{\dot M}{2\pi} \left( \Omega - \frac{\lambda}{r^2}
       \right)~.
\end{equation}
        Here, $\Omega \equiv V_\phi / r$ is the angular
velocity of the disk,
and
$\lambda \equiv \dot L / \dot M >0$ is
an angular momentum per unit mass
 of the accretion flow.
        The  viscous stresses are typically very small
at the inner radius of the disk where  the motion departs
significantly from Keplerian.
       The inner radius of the disk
determined by this way is $ r_i \approx{\lambda^2}/{GM}$.
       The ``viscous'' stresses are
determined by  processes of turbulent fluctuations  which are not
fully understood.
         A simplified
approach is commonly adopted where the $r\phi$-component of the
viscous stress tensor (integrated over $z$) is expressed as $
\tau=-\nu_t \Sigma r {\partial \Omega}/{\partial r}$, where $\nu_t$
is coefficient of turbulent viscosity.
        In the region of
quasi-Keplerian disk $(\Omega \approx \Omega_K)$
$$
\tau= \frac{3}{2} \nu_t \Sigma \Omega_K~,
$$
\begin{equation}\label{eq3}
V_r=-\frac{3\nu_t}{2r}~,
\end{equation}
$$
\Sigma=\frac{\dot{M}}{3\pi \nu_t}~.
$$
      Three-dimensional simulations of the local
regions of Keplerian disks indicate that the viscous stress $\tau$
constitutes  a fraction of the gas pressure (integrated over $z$) in
the disk.
         Namely, $\tau=({3\alpha_{\rm v}}/{2}) \Pi = ({3\alpha_{\rm v}}/{2}) \int
dz~p $, where $\alpha_{\rm v}$ is a dimensionless coefficient.
      The value of $\alpha_{\rm v}$ is in
the rough interval $3 \times 10^{-3} - 0.4$ (Balbus 2003).
        Comparing the
last expression and (\ref{eq3}), we obtain the Shakura-Sunyaev
formulae $\displaystyle{\nu_t={\alpha_{\rm v} c_s^2}/{\Omega_K}}$,
where $\displaystyle{c_s^2=\Pi/ \Sigma}$ is an average isothermal
speed in the disk.

         Equations  (\ref{eq3}) show that the
radial accretion speed  is determined by local properties of the
disk.
         The  surface density $\Sigma$ is set
by the local properties as well as the global quantity, the
accretion rate $\dot{M}$.
        The same is true for the nominal inner radius
of the disk $r_i$, which is determined by the parameter $\lambda$.
        To understand
      how the processes
at the inner edge of the disk influence the global characteristics
of accretion process $\dot{M}$ and $\dot{L}$, it is necessary to
consider non-stationary processes.
         The continuity equation gives
\begin{equation}\label{eq4}
\frac{\partial \Sigma}{\partial t} + \frac{1}{r}
\frac{\partial}{\partial r}\big( r \Sigma V_r \big) =0~.
\end{equation}
Conservation of angular momentum gives
\begin{equation}\label{eq5}
\frac{\partial (\Sigma r V_\phi)}{\partial t} + \frac{1}{r}
\frac{\partial}{\partial r} \big[ r^2 \left( \Sigma V_rV_{\phi}
+\tau\right)\big] =0~.
\end{equation}
If $\Omega\approx\Omega_K$ we obtain
\begin{equation}\label{eq6}
r \Sigma V_r = - \frac{1}{d\ell/dr} \frac{\partial
(r^2\tau)}{\partial r} = - \frac{2}{V_K} \frac{\partial( r^2
\tau)}{\partial r}~,
\end{equation}
where $\ell= \Omega_Kr^2$  is specific angular momentum.
Taking 
 the first of equations (3),
$$
\Sigma=\frac{2}{3} \frac{\tau}{\nu_t \Omega_K}~~~,
$$
and combining
this equation with  (\ref{eq4}), and (\ref{eq6})
we obtain
\begin{equation}
\frac{\partial f}{\partial t} = \frac{3 \nu_t}{\sqrt{r}}
\frac{\partial}{\partial r} \left( \sqrt{r} \frac{\partial
f}{\partial r} \right)~,
\end{equation}
where $f\equiv r^2 \tau$ is the angular momentum flux due to the
viscous stress.

       The evolution of $f=r^2\tau$ is described
by a diffusion equation with diffusion coefficient  $\sim \nu_t$.
      Thus, perturbations generated
for example by the star's magnetic field at the inner radius of the
disk with periodicity $\Delta t$ diffuse out to distances
$\displaystyle{\Delta r \sim \sqrt{\nu_t \Delta t} \sim c_s
\sqrt{{\alpha_{\rm v} \Delta t}/{\Omega_K}}}$. If
$\displaystyle{\Delta t \sim 1/ \Omega_K}$, then
$\displaystyle{\Delta r \sim {c_s} \sqrt{\alpha_{\rm v}}/{\Omega_K}
\leq h}$.
       Thus, disturbances of the inner part  of the
disk at say $r_d$ are not expected to influence the more distant
parts of the disk at $r\gtrsim 2 r_d$.

\subsection{Analytic Model of
Disk Accretion in ``Propeller'' Regime}

Here, we consider the influence on the disk of an aligned dipole
magnetic field of a rotating star following the treatment of
Lovelace, Romanova, and Bisnovatyi-Kogan
1999(LRBK99).

\begin{figure*}[t]
\epsscale{0.9} \plotone{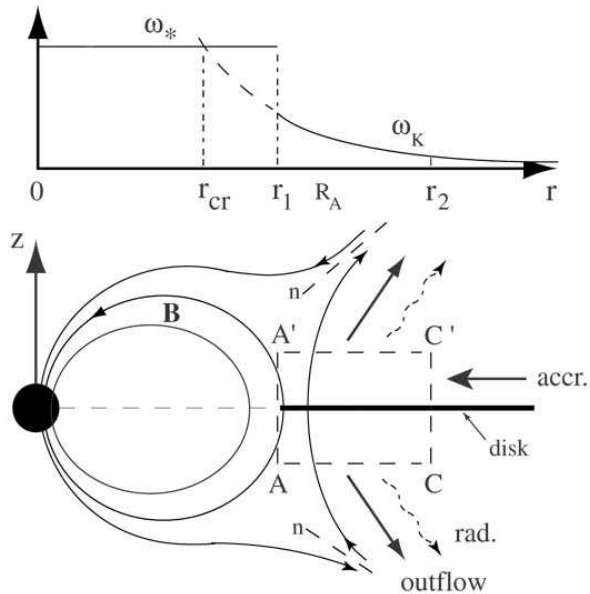} \caption{Geometry of disk accretion
to a rapidly rotating star with an aligned dipole magnetic field.
Top panel shows radial distribution of angular velocities of the
star and of the disk. Bottom panel shows schematic structure of the
magnetosphere in the propeller regime (from Lovelace et al. 1999). }
\label{r0}
\end{figure*}

       The balance of mass, angular momentum
and energy flows is considered in the region $AA'C'C$ of Figure
\ref{r0}, where the surface $AA'$ is along the inner region of the
disk, while the surface $CC'$ is sufficiently far away so that the
flow can be considered to be undisturbed by the magnetic field of
the star.
     The star's mass is $M$, magnetic moment $\mu$, and
angular velocity $\Omega_*$.
       The accretion disk with accretion rate $\dot{M}$ is
located at $r\geq r_d$
and
 rotates approximately with Keplerian velocity.
        The magnetosphere rotates with angular
velocity of the star $\Omega_*$.
       It is accepted that the angular
momentum flux through the surface $AA'$ is $ N={\alpha
\mu^2}/{r^3} $, where $\alpha$ is a dimensionless constant which
is an analogue of $\alpha$-coefficient in the model by
Shakura-Sunyaev.
        The inner
radius of the disk $r_d$ is found to be
\begin{equation}\label{eq8}
\kappa \left( \frac{r_A}{r_d} \right)^{7/3} = \frac{2\alpha}{3}
\frac{r_A^{7/2}}{r_d^2r_c^{3/2}} - 1 ~.
\end{equation}
Here,  $r_c=( GM / \Omega_*^2 )^{1/3}$ is the corotation radius,
$r_A= ( \mu^4 /GM \dot{M}^2 )^{1/7}$ is the nominal Alfv\'en
radius, $\kappa \lesssim 1$ is a dimensionless constant. (We have
simplified the
LRBK99
 treatment by taking $\delta=0$ in original formulae.)

      Equation (\ref{eq8}) was obtained
under fairly general assumptions. However,  the derivation assumed
stationary conditions and a specific expression for the wind mass,
energy, and angular momentum outflow was adopted from Lovelace,
Berk, and Contopoulos (1991).
      Equation (\ref{eq8}) implies that
the inner radius of  the disk $r_d$ or effective Alfv\'en radius is
determined not only by the mass and magnetic moment of the star and
the accretion rate, but also by the star's angular velocity.

       The processes giving the ``friction'' at
the boundary between the inner edge of the disk and magnetosphere is
treated phenomenologically by
LRBK99
in terms of the dimensionless coefficient $\alpha$.
        The
process
         evidently
has a turbulent nature and needs a separate detailed investigation
(see, e.g., Rast\"atter \& Schindler 1999). 

\subsection{Disk with Finite Conductivity}

       The interaction between the
magnetized star and the disk depends significantly on the conditions
in the corona.
       Magnetic flux of the star extends into the
corona and may thread the disk.
       We consider the corona to be a perfect conductor.
        Differential rotation of the
foot-points of magnetic field loops threading the star and the disk
tend to inflate and open unless there are some opposing factors
which suppress this tendency (e.g., Gold \& Hoyle 1960; Aly 1980;
Lovelace, Romanova \& Bisnovatyi-Kogan 1995; 
Uzdensky 2002). This was also confirmed by a number of numerical
simulations (e.g., Goodson, B\"ohm \& Winglee 1999;  Fendt \&
Elstner 2002).

One factor is the finite conductivity
       of
the disk plasma
which leads to reconnection of the inflated field lines (e.g., Aly
\& Kuijpers 1990; Hayashi, Shibata \& Matsumoto 1996; Uzdensky,
K\"onigl \& Litwin 2002; Yelenina \& Ustyugova 2005).
 A second factor is the density of the corona.
A sufficiently dense corona acts to inhibit opening of the  magnetic
field.
        Alternatively, a
rarefied corona favors the opening of field lines with the coronal
field passing through a sequence of force-free configurations.
        The finite conductivity of the
disk leads to the slippage of magnetic field lines relative to the
disk matter and thus limits the build up of the toroidal magnetic
field.
      A dense corona acts to oppose
twisting of
 the magnetic field.

       We first consider the influence of
the finite conductivity of the disk which is assumed to be thin.
        In a coordinate system attached to
a point in the disk Ohm's law is satisfied in the form ${\bf i} =
\lambda {\bf E}_t$, where $\bf i$ is the surface current density,
${\bf E}_t$ is the
electric field in the plane of the disk, and  $\lambda$ is the
``surface  conductivity'' ($\int dz \sigma$). Thus, in  the corona
in the non-rotating frame, ${\bf E}=-{\bf v}\times {\bf B}/c$, where
${\bf v}$ is the flow velocity in the corona. In the disk, ${\bf
i}=\lambda({\bf E}+{\bf V}\times {\bf B}/c)$, where ${\bf V}$ is the
flow velocity in the disk. These relations can readily be combined
to give
\begin{equation}\label{eq9}
\displaystyle{[ ({\bf v}-{\bf V}+ \zeta~ \hat{\bf z})\times{\bf
B}]_t=0}~,
\end{equation}
on the top side of the disk.
      Here, $\zeta\equiv c^2 /(2\pi \lambda)$ is the surface
magnetic diffusivity, $\hat{\bf z}$ is the unit normal to the disk,
and index $t$ shows that at the left-hand-side only tangential to
the disk components of vector are important.
      If we suppose that there is no outflow from the disk
     ($v_z=0$), then it follows from
(\ref{eq9}) that a toroidal magnetic field is generated at its
surface:
\begin{equation}\label{eq10}
\displaystyle{B_\phi=\frac{v_\phi-V_\phi}{\zeta} B_z}~~~.
\end{equation}

       In order to calculate the  toroidal magnetic
field we make a number of assumptions:
       First, for the coefficient of magnetic
diffusivity we follow the Shakura and Sunyaev $\alpha$ prescription,
$\eta_t=\alpha_{\rm d} c_s h$, here $c_s$ is an isothermal sound
speed (Bisnovatyi-Kogan \& Ruzmaiken 1974).
       Second,
the magnetic field in the disk $B_z$ is taken to be the unperturbed
dipole  field of the star, $B_z=\mu /r^3$.
       Thirdly, the disk is assumed to rotate with Keplerian
angular velocity while the coronal matter above the disk rotates
with the angular velocity of the star, $\Omega_*$. Neglecting the
variations of the tangential electric field ${\bf E}_t$ across the
disk, we obtain
\begin{equation}\label{eq11}
\frac{1}{\zeta} = \int_{-h}^h \frac{dz}{\eta_t} =
\frac{2}{\alpha_{\rm d} c_s} ~.
\end{equation}
Substituting  (\ref{eq11}) to (\ref{eq10}) we obtain
$$
B_\phi = \frac{2r}{\alpha_{\rm d} c_s} (\Omega_*-\Omega_K) B_z =
\frac{2\mu}{\alpha_{\rm d} c_s r^2} (\Omega_*-\Omega_K)~.
$$
For $\alpha_{\rm d} \ll 1$ there is a strong twisting of the
magnetic field linking the star and the disk,
$$ \left|
\frac{B_\phi}{B_z} \right| = \frac{2r
|\Omega_*-\Omega_K|}{\alpha_{\rm d} c_s} \gg 1~,
$$
if the considered distance $r$ is far from the corotation radius
where $\Omega_*=\Omega_K$.
      This equation was derived earlier by Lovelace et al. (1991).

      The discontinuity of the azimuthal
component of the magnetic field at the disk surface $z=0$ means that
there is a radial surface current flowing along the disk.
      The interaction of this current
with the vertical magnetic field, $B_z$, leads to the loss of
angular momentum from both sides of the disk at the rate $-2 r
B_\phi B_z /4\pi$ per unit surface area.
        For stationary conditions, the conservation
of the angular momentum balance gives
$$
\frac{1}{r} \frac{\partial}{\partial r}\left[ r^2 \left( \Sigma V_r
V_\phi +\tau \right)\right] = \frac{rB_\phi B_z}{2\pi}=
\frac{\mu^2}{\pi \alpha_{\rm d}c_s r^4} (\Omega_*-\Omega_K)~.
$$
Assuming $\alpha_{\rm d},~ c_s ={\rm const}$, and $V_\phi=V_K(r)$,
this equation can be integrated to give
\begin{equation}\label{eq12}
\tau=\frac{\dot{M}}{2\pi} \left(\Omega_K-\frac{\lambda}{r^2} \right)
+ \frac{\mu^2}{\pi \alpha_{\rm d} c_s} \left(\frac{2
\sqrt{GM}}{7r^{11/2}} - \frac{\Omega_*}{2r^4} \right)~~~,
\end{equation}
where $\lambda  = \dot{L}/\dot{M} $ is a constant of integration.

We identify the point $r=r_d$, where $\tau=0$, with the inner radius
of the disk.
      At this distance
the  viscous and magnetic stresses lead to significant deviation of
rotational velocity from Keplerian.
         For a sufficiently
large magnetic moment and a rapidly rotating star, the deviation
from Keplerian motion is determined by the magnetic field.
       We can then estimate $r_d$ by
comparing the first and the last terms on the right-hand side
(\ref{eq12}).
       This gives
$$
r_d^{5/2}=\frac{\mu^2\Omega_*}{\dot{M}\sqrt{GM}\alpha_{\rm d} c_s}
=\frac{r_A^{7/2}}{\alpha_{\rm d} h_*}~,
$$
     where $r_A=( \mu^4 /GM\dot{M}^2)^{1/7}$
is the nominal Alfv\'en radius and $h_*=c_s/\Omega_*$ is the
hydrostatic thickness of the disk at the corotation radius.
       Thus, $r_d=r_A ( r_A /\alpha_{\rm d} h_*)^{2/5} \gg r_A $.
This derivation suggests that $r_d\gg r_{c}$ and $r_d\gg
r_i=\lambda^2 /GM$.

       In order to estimate the influence of
the magnetic field on the disk, we  consider  a tube of field lines
with one foot point frozen to the star and rotating with the angular
velocity $\Omega_*$, and the other foot-point in the disk rotating
with angular velocity $\Omega \neq \Omega_*$.
       The field lines
will be deformed due to the differential rotation and this gives
rise to an azimuthal magnetic field component.
        Furthermore, the field lines tend to
open as a result of the differential rotation.
       At the same time
the azimuthal component of the field acts in combination with the
poloidal field applies a  torque to the disk matter.
       To estimate the net effect on the disk
we assume that the opening of the field lines occurs on a time scale
$ \Delta t \sim 1/\Delta \Omega$, where $\Delta \Omega =
|\Omega_{*}-\Omega|$.
      During this time the azimuthal
magnetic field builds up to a value $B_\phi \sim B_z$ at the surface
of the disk.
      The magnetic stress on the disk
during this process exerts a torque on the disk (from both sides).
       The corresponding loss of angular momentum (per
unit area) from both sides of the disk is $\Delta t ~r B_\phi B_z
/2\pi
      \sim r B_z^2 /(2\pi \Delta \Omega)$.
      Of course angular momentum per unit
area of the disk is simply $ r^2 \Sigma \Omega$.
        The magnetic field will
have a significant influence on the disk when
$$
\frac{r B_z^2}{2 \pi \Delta \Omega} \gtrsim r^2 \Sigma \Omega~.
$$
         Assuming the disk to be roughly Keplerian, $\Omega
\approx \sqrt{GM /r^3}$ with $\Sigma=\dot{M} /3\pi \nu_t$, and  the
magnetic field  $B_z=\mu /r^3$, we obtain the condition
\begin{equation}
\label{eq13} \Omega |\Omega_*-\Omega| \lesssim \frac{\mu^2/r^7}{2
\pi \dot{M}/(3\pi \nu_t)} = \frac{3 \nu_t}{2} \frac{\mu^2}{\dot{M}
r^7}~,
\end{equation}
for the magnetic field to influence the disk motion.

\begin{figure*}[t]
\epsscale{1.1} \plotone{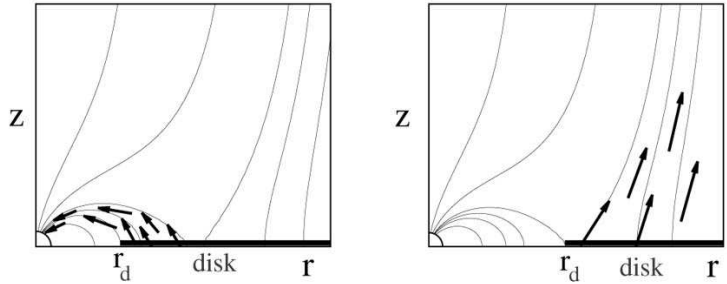} \caption{Sketches  of matter flow in
the non-propeller (left panel) and propeller (right panel) cases.}
\label{r1}
\end{figure*}

\begin{figure*}[t]
\epsscale{1.1} \plotone{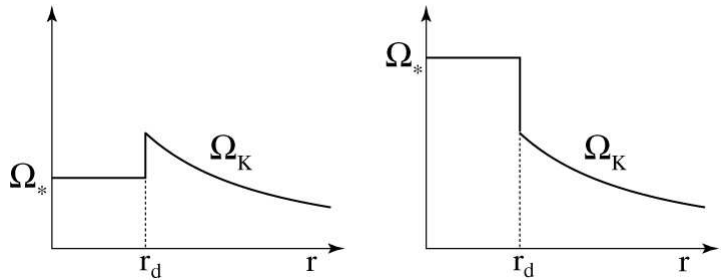} \caption{Sketches of the radial
distribution of the angular velocity in the non-propeller (left
panel) and the propeller (right panel) cases.} \label{r2}
\end{figure*}

      In the limits of slowly and rapidly rotating
stars, inequality (\ref{eq13}) provides an estimate of the inner
radius of the  disk:
        For a slowly rotating star,
$(\Omega_* \ll \Omega)$,
$$
       r_{d} \sim \left( \frac{\alpha_{\rm v}^2 c_s^4 \mu^4}
{G^3M^3\dot{M}^2}\right)^{1/7}
       = r_A \left( \frac{\alpha_{\rm v}^2 c^4}{G^2M^2} \right)^{1/7}~.
$$
For a rapidly rotating star, $(\Omega_* \gg \Omega)$,
$$
       r_{d} \sim \left( \frac{\alpha_{\rm v} c^2 \mu^2}
{GM\dot{M}\Omega_*} \right)^{1/4} ~.
$$
Here, we have assumed $\nu_t=\alpha_{\rm v} c_s^2/\Omega_K$,
$\alpha_{\rm v}={\rm const }$ and dropped insignificant numerical
coefficients.
      The first case corresponds to
the situation where the corotation radius $r_c$ is much larger than
the inner disk radius.
       The second case
corresponds to the opposite inequality.
      In the first case the
disruption of the disk occurs in the {\it Funnel Flow} regime.
      Matter loses its angular momentum because of ``friction''
with the slowly rotating magnetosphere and accretes to the star
along the dipole field lines, forming the funnel flows. In the
second case the disruption of the disk occurs in the {\it Propeller
Regime} where matter acquires angular momentum ``friction'' with the
rotating magnetosphere and is driven  away from the star, forming
the outflows (jets and winds).
       Sketches of the flows  are shown in
the figure \ref{r1}.
      Sketches of the radial variation of the disk
angular velocity in the equatorial plane in the two cases are shown
in Figure \ref{r2}.

\begin{figure*}[t]
\epsscale{1.1} \plotone{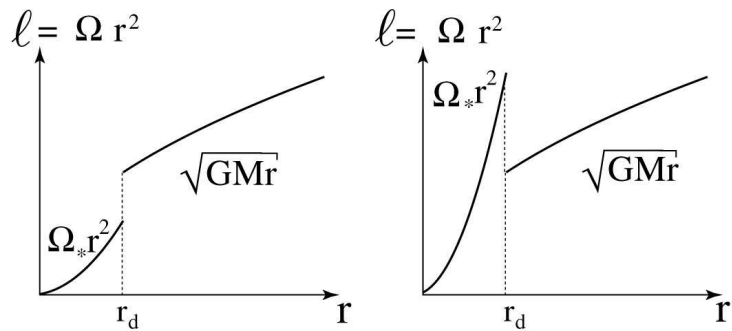} \caption{Sketches of the radial
distribution of the specific angular momentum in the non-propeller
(left panel) and propeller (right panel) cases.} \label{r3}
\end{figure*}

      The transition at $\sim r_d$ between the magnetosphere
region which rotates with the angular velocity of the star
$\Omega_*$ and the disk which rotates with Keplerian angular
velocity $\Omega_K$ (more exactly a part of the disk which rotates
with quasi-Keplerian velocity) is not sharp, but instead occurs
over some distance where instabilities may have a significant
role.
         Instabilities can act generate turbulence
at the inner edge of the disk. We expect turbulence will be strongly
developed in the ``propeller'' regime. Figure \ref{r3} shows the
dependence of the specific angular momentum $\ell=\Omega r^2$ on
radius in the two cases.

One can see that in the ``propeller'' regime there is a region where
$\partial \ell /\partial r <0$, which is the Rayleigh criterion for
axisymmetric instability (Chandrasekhar 1961).
      The decrease of
the angular momentum with the distance is only one of the reasons
which may give rise to turbulence in the disk.
       Nevertheless, it is reasonable to expect
that the turbulent viscosity and magnetic diffusivity are
significantly enhanced at the inner edge of the disk compared with
the regions of the Keplerian disk at larger distances.

\section{Formulation of the Problem}

\subsection{Physics Problem}

       As discussed the nature of the
interaction of the magnetized star with the accretion disk depends
on many factors.
      Nevertheless, we can try to classify the
resulting configurations basing on the relations between three
lengths, the corotation radius $r_{c}$, the magnetospheric radius
$r_m$, and the inner radius of the disk $r_d$.
       We recall that for
$r_m$ we have in mind the location  in the equatorial plane where
\begin{equation}
\label{eq14} p+\rho v_\phi^2=\frac{{\bf B}^2}{8\pi}~.
\end{equation}
        It is convenient to use the radius $r_m$
for comparison with the other two radii even though it depends on
the results of simulations.
       We also
note that in the subsequent parts of the paper we will consider only
the ``propeller" regime.
       That is,  the star is rapidly rotating
in the sense that $r_c < r_d$.
       Classification based on the ratio between $r_d$ and
$r_m$ is not similarly clear, because all variables in the
(\ref{eq14}) are rapidly varying functions of the position of the
point.
       This is why significant variation
of the parameters of the system, for example,  the magnetic moment
of the star, may not lead to significant variation of $r_m$.
       An important quantity
is sound speed in the corona, which determines variations of the
density and pressure (see section on Initial conditions).
       Results of our numerical simulations show
that it is always a true that $r_d \approx r_m$, but in spite of
that the matter flow may have different morphology.
      In a number of cases the magnetic field forms
configuration in which some of the poloidal field lines are
well-collimated along the symmetry axis, while at larger distances
the field part is pushed toward the equatorial plane  (RUKL04).
       In other cases the collimation along the open
field lines is less prominent, but matter flows from the disk along
the neutral layer where
${\bf B}_p \approx 0$ (see also RUKL05).
       We trace such qualitative
differences in the final picture of the disk-star interaction to the
initial temperature or sound speed in the corona which determines
the initial differences of density and pressure.

In the considered model we assume a  cold quasi-Keplerian disk
surrounded by a relatively hot corona. The disk and corona are in
the gravitational field of the central object, a star of mass $M$
which rotates with angular velocity $\Omega_*$ around the same axis
as the disk.
        The star has a
magnetic moment $\rvecmu$ which is aligned with the rotational axis
of the star.

Having in mind application of the results of the numerical
modeling to T Tauri type stars, we mention typical parameters, the
mass of the star, $M=0.8 M_{\odot} = 1.6\times 10^{33}$g, its
radius $r_*=2 r_{\odot} = 1.4 \times 10^{11}$ cm, the inner radius
of the disk, $r_d=5r_*=7 \times 10^{11}$ cm, the magnetic field at
the surface of the star (in the equatorial plane), $B_*=10^3$G,
the temperature of the corona $T_{cor}=10^6K$, and the temperature
of the disk $T_{disk}=5 \times 10^3K$.

The Keplerian velocity at the inner radius of the disk is
$v_K=\sqrt{GM /r_d} \approx 1.2 \times 10^7$ cm/s.
       The  isothermal sound speed in the corona is
$c_{cor}=\sqrt{RT_{cor}/m} \approx 1.3 \times 10^7$ cm/s (for fully
ionized hydrogen $m=1/2$); in the disk $c_{disk}=\sqrt{RT_{disk}} =
6.5 \times 10^5$ cm/s, where $R=8.31\times 10^7$ergs/(mol K) is the
gaseous constant.
      The magnetic moment of the star is
$\mu=B_*r_*^3=3.43\times 10^{38}$G cm$^3$.
       We treat the density in both the disk
and the corona as free parameters.
      For a fixed sound speed in
the corona (or the disk) and a given magnetic field of the star, the
choice of say the disk density  determines the ratio of the Alfv\'en
speed and  the sound speed, or equivalently  the ratio of the
magnetic and gas pressures.
       We  chose  the density  such that  in at least
part of corona near the star, the magnetic energy density dominates,
${\bf B}^2 /8 \pi > \rho v_\phi^2+p$.
        At the inner edge of the disk the
Keplerian velocity is related to the sound speed as
$v_K:c_{cor}:c_{disk} \approx 1:1:0.05$.

\subsection{Basic Equations}

       We assume that the plasma flows are described by
the equations of magnetohydrodynamics.
      Further, we consider that  any shock waves
in the flow are outside of the considered region, for example,
inside the considered region, on the surface of the star.
        This allows us to use an  energy
conservation equation in the form of an entropy continuity equation.
The system of equations in  a non-rotating reference
frame is:
\begin{equation}\label{eq15}
\begin{array}{rcl}
\displaystyle{ \frac{\partial \rho}{\partial t} + {\bf \nabla}\cdot
\left( \rho
{\bf v} \right)}& = &0~,\\[0.5cm]
\displaystyle{\frac{\partial \rho {\bf v}}{\partial t} + {\bf
\nabla}\cdot
{\cal T} }& = &\rho ~{\bf g}~,\\[0.5cm]
\displaystyle{\frac{\partial {\bf B}}{\partial t} + c {\bf
\nabla}\times
{\bf E}} & = &0~,\\[0.5cm]
\displaystyle{\frac{\partial (\rho S)}{\partial t} + {\bf
\nabla}\cdot ( \rho S {\bf v} )}& = & Q~.\
\end{array}
\end{equation}
Here, $\rho$ is the density and $S$ is the specific entropy; $\bf v$
is the flow velocity; $\bf B$ the magnetic field; $\cal{T}$ the
momentum flux-density tensor; $\bf E$ the electric field; $Q$ is the
rate of production or loss of  entropy per unit volume; $c$ is speed
of light;  and ${\bf g} = - (GM /r^2)\hat{{\bf r}}$ is the
gravitational acceleration due to the star of mass $M$. The total
mass of the disk is negligible compared to $M$.

      We consider axisymmetric MHD-flows
and use spherical coordinates $(r, \theta, \phi)$. The polar angle
counted from symmetry axis, all derivatives $\partial /\partial
\phi =0$. In spherical coordinates equations (\ref{eq15}) have the
form:
$$
\begin{array}{rcl}
\displaystyle{ \frac{\partial \rho }{\partial t} + \frac{1}{r^2}
\frac{\partial}{\partial r}\left( r^2 \rho v_r\right) +
\frac{1}{r\sin\theta} \frac{\partial}{\partial \theta} \big( \sin
\theta \rho v_{\theta}\big) }& = &
0~,\\[0.8cm]
\displaystyle{ \frac{\partial ( \rho v_r) }{\partial t} +
\frac{1}{r^2} \frac{\partial}{\partial r} \left( r^2 {\cal
T}_{rr}\right) + \frac{1}{r \sin \theta} \frac{\partial}{\partial
\theta} \big(\sin
\theta {\cal T}_{r \theta}\big) }& = &\\[0.8cm]
= \displaystyle{\frac{{\cal T}_{\theta \theta}+
{\cal T}_{\phi \phi}}{r}+\rho g_r~,}&&\\[0.8cm]
\displaystyle{\frac{\partial (\rho v_{\phi})}{\partial t} +
\frac{1}{r^3} \frac{\partial}{\partial r} \big( r^3 {\cal T}_{r
\phi}\big) + \frac{1}{r \sin^2 \theta} \frac{\partial}{\partial
\theta} \big(\sin^2
\theta {\cal T}_{\theta \phi}\big) }& = & 0~,\\[0.8cm]
\displaystyle{\frac{\partial (\rho v_{\theta})}{\partial t} +
\frac{1}{r^3} \frac{\partial}{\partial r} \big(r^3 {\cal T}_{r
\theta}\big) + \frac{1}{r} \frac{\partial}{\partial \theta} {\cal
T}_{\theta
\theta} }& = & \\[0.8cm]
= \displaystyle{{\rm cotan} \theta \frac{{\cal T}_{\phi \phi}-
{\cal T}_{\theta \theta}}{r}~,}&&\\[0.8cm]
\displaystyle{\frac{\partial B_r }{\partial t} + \frac{1}{r \sin
\theta} \frac{\partial}{\partial \theta} \big(\sin \theta c
E_{\phi}\big)} & = &
0~,\\[0.8cm]
\displaystyle{\frac{\partial B_{\phi} }{\partial t} + \frac{1}{r}
\frac{\partial}{\partial r} \left( rcE_{\theta} \right)- \frac{1}{r}
\frac{\partial}{\partial \theta} \big(cE_r\big)}& = &
0~,\\[0.8cm]
\displaystyle{\frac{\partial B_{\theta} }{\partial t} + \frac{1}{r}
\frac{\partial}{\partial r} \left( -cE_{\phi} \right)}& = &
0~,\\[0.8cm]
\displaystyle{\frac{\partial (\rho S)}{\partial t} + \frac{1}{r^2}
\frac{\partial}{\partial r} r^2 \big(\rho S v_r\big) + \frac{1}{r
\sin \theta} \frac{\partial}{\partial \theta}\big( \sin \theta \rho
S v_{\theta}\big) }& = & Q~~~.
\end{array}
$$
        The stress tensor $\cal T$ is
$$
\displaystyle{{\cal T}_{ik}=\rho v_i v_k +p \delta_{ik} + \left(
\frac{B^2}{8 \pi} \delta_{ik} - \frac{B_iB_k}{4 \pi} \right) +
\tau_{ik}= T_{ik}+\tau_{ik}}~.
$$
Here, $p$ is the pressure; $\tau_{ik}$ is the ``viscous" stress
caused by the turbulent fluctuations of the velocity and magnetic
field; $g_r=-GM /r^2$ is the acceleration due to gravity.
        The  plasma is considered to be an
ideal gas with adiabatic index $\gamma =5/3$, so that $S=\ln(p/
\rho^{\gamma})$.

We assume that the stress due to fluctuations of the velocity and
the magnetic field can be represented in the same way as collisional
viscosity by substitution of the  turbulent viscosity coefficient.
      Moreover, we consider that the viscous stress
is determined mainly by the gradient of the angular velocity because
the azimuthal velocity is the dominant component in the disk.
        The dominant components of the tensor
$\tau_{ik}$ in spherical coordinates are:
$$
\tau_{r \phi} = - \nu_t \rho r \sin \theta \frac{\partial
\omega}{\partial r}~,
$$
$$
\tau_{\theta \phi} = - \nu_t \rho \sin \theta \frac{\partial
\omega}{\partial \theta}~.
$$
Here $\omega = v_{\phi} /r \sin \theta$ is angular velocity of the
plasma and $\nu_t$ is the coefficient of the turbulent viscosity.

      Separating out the viscous stress gives
$$
\frac{\partial (\rho v_{\phi})}{\partial t} + \frac{1}{r^3}
\frac{\partial( r^3 T_{r \phi})}{\partial r} + \frac{1}{r \sin^2
\theta} \frac{\partial (\sin^2 \theta T_{\theta \phi})}{\partial
\theta} =
$$
\begin{equation}
\label{eq16}
     \frac{1}{r^3} \frac{\partial}{\partial r} \left( \nu_t \rho
r^4 \sin \theta \frac{\partial \omega}{\partial r} \right) +
\frac{1}{r \sin^2 \theta} \frac{\partial}{\partial \theta} \left(
\nu_t \rho \sin^3 \theta \frac{\partial \omega}{\partial \theta}
\right) ~,
\end{equation}
where $T_{r \phi}$ and $T_{\theta \phi}$ are components of the
inviscid part of the stress tensor.

       The viscosity leads of course to dissipation of the
kinetic energy and its conversion to thermal energy and to a
corresponding increase of the entropy. The rate of energy
dissipation per unit volume is
$$
\frac{\tau_{ik}^2}{2 \rho \nu_t} = \frac{\rho \nu_t}{2} \sin^2
\theta \left[ r^2 \left( \frac{\partial \omega}{\partial r}
\right)^2 + \left( \frac{\partial \omega}{\partial \theta} \right)^2
\right]~~~.
$$
We assume that the finite conductivity of plasma as well as
``viscous" stresses are connected mainly with the turbulent
fluctuations of the velocity and magnetic field.
      The induction equation averaged over
the small-scale fluctuations has the form
\begin{equation}
\label{eq17} \frac{\partial {\bf B}}{\partial t} - {\bf
\nabla}\times ({\bf v}\times{\bf B}) + c {\bf \nabla}\times {\bf
E}^\dagger =0~.
\end{equation}
Here, $\bf v$ and ${\bf B}$ are the averaged velocity and magnetic
fields, and ${\bf E}^\dagger=- \left< ({\bf v}'\times{\bf B}')
\right>/c$  is electromotive force connected with the fluctuating
fields.
      Because the turbulent electromotive
force ${\bf E}^\dagger$ is connected with the small-scale
fluctuations, it is reasonable to suppose that it has a simple
relation to the ordered magnetic field $\bf B$.
       If we neglect the magnetic dynamo
$\alpha$-effect (Moffat 1978), then $\left< ({\bf v}'\times{\bf B}')
\right> = -\eta_t {\bf \nabla}\times {\bf B}$, where $\eta_t$ is the
coefficient of  turbulent magnetic diffusivity. Equation
(\ref{eq17}) now takes the form
\begin{equation}
\label{eq18} \frac{\partial {\bf B}}{\partial t} - {\bf
\nabla}\times ({\bf v} \times {\bf B}) + {\bf \nabla} \times\left(
\eta_t {\bf \nabla}\times {\bf B} \right) =0~.
\end{equation}
We should note that the term for ${\bf E}^\dagger$ formally
coincides with the Ohm's law
$$
{\bf J} = \frac{c}{4\pi} {\bf \nabla}\times {\bf B} =
\frac{c^2}{4\pi \eta_t} {\bf E}^\dagger~.
$$
The coefficient of turbulent electric conductivity $\sigma = c^2
/4\pi \eta_t$. The rate of dissipation of  magnetic energy per unit
volume is
$$
\frac{{\bf J}^2}{\sigma} = \frac{\eta_t}{4 \pi} ({\bf \nabla}\times
{\bf B})^2~.
$$

To calculate the evolution of the poloidal magnetic field it is
useful to calculate the $\phi$-component of the vector-potential
$\bf A$.
       Owing to the assumed axisymmetry,
$$
B_r = \frac{1}{r \sin \theta} \frac{\partial (\sin \theta
A_{\phi})}{\partial \theta}~,\quad\quad B_{\theta} = - \frac{1}{r}
\frac{\partial (r A_{\phi})}{\partial r}~.
$$
Substituting  ${\bf B} = {\bf \nabla}\times {\bf A}$ to the
induction equation gives
\begin{equation}
\label{eq19} \frac{\partial {\bf A}}{\partial t} + c {\bf E} =
\nabla \chi~,
\end{equation}
where $\chi$ is an arbitrary  function.
       Thus,
\begin{equation}
\label{eq20} \frac{\partial A_{\phi}}{\partial t} - ({\bf
v}\times{\bf B})_{\phi} + \eta_t ({\bf \nabla}\times {\bf B})_\phi
=0~,
\end{equation}
or
$$
\frac{\partial A_{\phi}}{\partial t} - \eta_t \left( \frac{1}{r}
\frac{\partial^2 (r A_{\phi})}{\partial r^2} + \frac{1}{r^2}
\frac{\partial}{\partial \theta} \frac{1}{\sin \theta}
\frac{\partial (\sin \theta A_{\phi})}{\partial \theta} \right) =
[{\bf v}\times{\bf B}]_{\phi}~.
$$
The azimuthal component of the induction equation gives
$$
\frac{\partial B_{\phi}}{\partial t} - \frac{1}{r}
\frac{\partial}{\partial r} \left( \eta_t \frac{\partial (r
B_{\phi})}{\partial r} \right) - \frac{1}{r^2}
\frac{\partial}{\partial \theta} \left( \frac{\eta_t}{\sin \theta}
\frac{\partial (\sin \theta B_{\phi})}{\partial \theta} \right)
$$
\begin{equation}
\label{eq21} =\frac{1}{r} \left( \frac{\partial [r ({\bf
v}\times{\bf B})_{\theta}]}{\partial r} - \frac{\partial ({\bf
v}\times{\bf B})_r}{\partial \theta} \right)~.
\end{equation}
The Joule heating rate per unit volume is
$$
\frac{\eta_t}{4\pi} ({\bf \nabla} \times{\bf
B})^2=\frac{\eta_t}{4\pi r^2} \times
$$
$$  \left[  \left( \frac{\partial( r B_{\theta})}{\partial r} -
\frac{\partial B_r}{\partial \theta} \right)^2 +
\left(\frac{\partial (r B_{\phi})}{\partial r} \right)^2 +
\frac{1}{\sin^2 \theta} \left( \frac{\partial( \sin \theta
B_{\phi})}{\partial \theta} \right)^2 \right]. $$

\subsection{Initial Data}

       A variety of simulation runs were calculated
for a range of values of the star's
     magnetic moment and its angular velocity and
for different values of the viscosity and magnetic diffusivity.
      Changes in the magnetic moment of the star (at $t=0$)
of course changes the location of the surface where $p+\rho
v_\phi^2={\bf B}^2/8 \pi$.

       The initial conditions have a significant
role in giving a ``smooth" transition of the system to possible
stationary, quasi-stationary or a quasi-periodic regime.
       This role of the initial conditions
provides a measure of their validity:
       the  time-averaged location of the Alfv\'en surface (in the
equatorial plane) should not be appreciably different from the its
initial value.
        In the opposite case, a significant part of the
computational time can be spent  for establishment of more or less
equilibrium distribution of the matter and the field.

\begin{figure*}[t]
\epsscale{1.1} \plotone{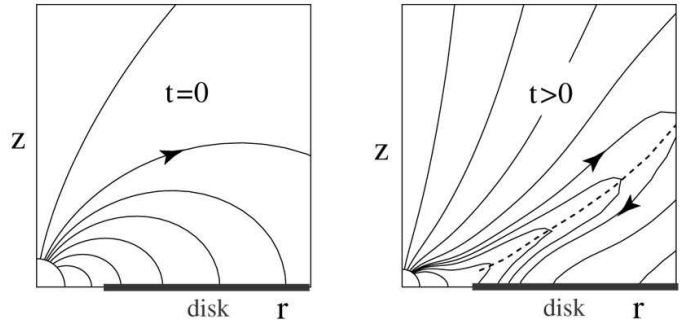} \caption{Sketches of the initial
(left) and final (right) configurations of the poloidal magnetic
field which results from the difference in angular velocities of the
star and the disk.} \label{r4}
\end{figure*}

         The initial
magnetic field is determined by the magnetic moment of the star.
      The field is a dipole with field lines
connecting the star and the disk.
       It is clear that the differential rotation of
matter along the field lines may lead to the generation of an
azimuthal magnetic field and to the opening of some of the field
lines (e.g., Lovelace et al. 1995).
       The qualitative
picture of the poloidal field lines  is shown in Figure \ref{r4}.
        The dashed line shows the line along which
the $r$-component of magnetic field is zero.
      A final field configuration could be
determined and used  as an initial condition.
         Nevertheless, we have found it efficacious
to start with  the poloidal dipole field, and let the system
evolution open the field lines
(see also Romanova et al. 2002).

       Initially the matter
of the disk and corona are assumed to be in mechanical
equilibrium, and the magnetic field is the current-free dipole
field of the star $B_r=2\mu \cos \theta /r^3$ and $B_\theta=\mu
\sin \theta /r^3$. In addition,  the initial density distribution
is taken to be barotropic with
$$
\rho (p)= \left \{
       \begin{array}{lcl}
         p/T_{disk}~,~~~ p>p_b~~~~ {\rm and}~~~~ r \sin \theta \geq
r_b \\[0.3cm]
         p/T_{cor}~,~~~~ p<p_b ~~~~{\rm or}~~~~ r \sin \theta \leq r_b~.
       \end{array}
\right.
$$
The level surface of pressure $p_b$ separates the cold matter of the
disk from the hot matter of corona. At this surface the density has
a discontinuity from value $p_b/T_{disk}$ to value $p_b/T_{cor}$.

       Because the density distribution is barotropic,
the angular velocity is a constant on coaxial cylinderical surfaces
about the $z-$axis.
         Consequently,
the pressure may be determined from
$$
F(p) + \Phi + \Phi_c =E = {\rm const}~.
$$
Here, $\Phi = -GM /|{\bf r}|$  is gravitational potential, $\Phi_c
  = \int_{r \sin \theta}^\infty \omega^2 (\xi)\xi
d\xi$ is centrifugal potential, which depends only on cylindrical
radius $ r\sin \theta$, and
$$
F(p)=\left \{
       \begin{array}{lcl}
         T_{disk} \ln( p/p_b )~,~~~ p>p_b~~~~{\rm and}~~~r \sin \theta >r_b~,
\\[0.3cm]
         T_{cor} \ln (p/p_b) ~,~~~~ p<p_b~~~~{\rm or}~~~~r\sin \theta <r_b~.
       \end{array}
\right.
$$

Initially, the inner edge of  the disk is located at $r_b=5$ in the
equatorial plane.
       The angular velocity of the disk is
slightly
sub-Keplerian $\Omega(\theta=\pi/2)=(1.-0.003)\Omega_K$, so that the
density and pressure decrease towards periphery.
       Inside the cylinder
$r\leq r_b$ the matter rotates rigidly with the angular velocity
$\Omega(r_b)=(1.-0.003)\sqrt{GM /r_b^3}$.
To insure the smooth start-up,
 the angular velocity of the star is changed gradually from its
initial value $\Omega_{cor} = 5^{-3/2} \approx 0.09$ to a final
value $\Omega_*$ during
three Keplerian rotation periods at $r=1$.

Below we show results of our simulations
for the following set of dimensionless parameters:
      the gravitational potential
is such that $GM=1$; the dipole moment of the star is $\mu=10$;  the
angular velocity of the star is $\Omega_*=1$; the initial
``temperature" of the disk is  $T_{disk}=\left( p /\rho
\right)_{disk}=0.0005$; the initial temperature of the corona is
     $T_{cor}=\left( p/\rho \right)_{cor}=0.5$.
      We  note that the dimensional temperature is
$T(K)=mv_0^2 T /R$, where $v_0$ is the velocity scale,  $m$ is the
number of barions per particle (including free electrons), which
by itself depends on the temperature, chemical composition and
other factors.

\subsection{Transition to the Dimensional Variables}

Results obtained in  dimensionless form may be applied to objects
with widely different scales.
     Below we give examples for classical T Tauri
stars (CTTSs) and for neutron stars with a weak magnetic field
(millisecond pulsars).

For CTTSs we adopt the following scales.
  The mass of the star is
$M_0=M_*=0.8 M_\odot=1.6\times10^{33}~{\rm g}$.
The length scale is
$r_0=2 r_*=4 r_\odot=2.8\times 10^{11}~{\rm cm}$.
   The magnetic field at
the surface of the star is $B_*=10^3~{\rm G}$.
   From these numbers
we obtain the other scaling variables: The time-scale
$t_0=0.166~{\rm days}$; velocity $v_0=1.95\times 10^7~ {\rm cm/s}$;
density $\rho_0=4.1\times 10^{-13}~{\rm g/cm^3}$; pressure
$p_0=156~{\rm erg/cm^3}$; accretion rate $\dot M_0=6.27\times
10^{17}~{\rm g/s} = 9.84\times 10^{-9}~{\rm M_\odot/yr}$; angular
momentum flux $\dot L_0=3.43\times 10^{36} ~{\rm g cm^2/s^2}$. For
these parameters, the magnetic moment of the star is
$\mu_*=2.74\times 10^{36}~{\rm G cm^3}$; the star's rotation period
is $P_*=1.04~{\rm days}$, the initial temperature in the disk is
$T_d=2,300~{\rm K}$, and coronal temperature is $T_c=1.15\times
10^6~{\rm K}$. Subsequently, we give time in periods of rotation of
the star.

For a neutron star, we give an illustration of the scales. The
star's mass is $M_0=1.5 M_\odot=3.0\times10^{33}~{\rm g}$. The
length scale is $r_0=2 r_*= 2\times 10^6~{\rm cm}$.
  The magnetic field at the
surface of the star is $B_*= 10^9~{\rm G}$.
  Other scaling variables include the
time-scale $t_0=2\times 10^{-4}~{\rm s}$, the velocity $v_0=10^{10}~
{\rm cm/s}$, the density $\rho_0=1.56\times10^{-6}~{\rm g/cm^3}$,
the pressure $p_0= 1.56\times 10^{14}~{\rm erg/cm^3}$, the accretion
rate $\dot M_0=6.25\times 10^{16}~{\rm g/s} = 10^{-9}~{\rm
M_\odot/yr}$, and the angular momentum flux $\dot L_0=1.25\times
10^{33} ~{\rm g cm^2/s^2}$.
   For these parameters, the magnetic moment
of the star is $\mu_*= 10^{27}~{\rm G cm^3}$, the rotation period of
the star is $P_*=1.26\times 10^{-3}~{\rm s}$, the initial
temperature in the disk is $T_d=3\times 10^8~{\rm K}$, and the
coronal temperature is $T_c=3\times 10^{11}~{\rm K}$.

\subsection{Numerical Method}

For the numerical integration
of the MHD equations including
magnetic diffusivity and viscosity,
we used a method of
splitting of the different physical processes.
      Our simulation algorithm
has a number of blocks: (1) a hydrodynamic block in which we
calculated the dynamics of the plasma and magnetic field with
dissipative processes switched off;  (2) a blocks for the
diffusion of the poloidal and azimuthal components of the magnetic
field calculated for frozen values of the plasma velocity and
thermodynamic parameters (density and pressure); and a  block for
the calculation of viscous dissipation in which we took into
account only $r \phi$ and $\theta \phi$ components of the viscous
stress tensor.

\begin{figure*}[t]
\epsscale{1} \plotone{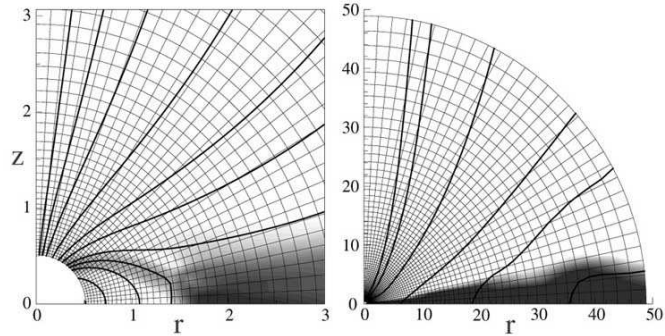} \caption{The left-hand panel shows the
inner part of the simulation region with the logarithm of density as
a background and the magnetic field lines (solid lines) and the
grid. The right-hand panel shows
the full simulation region.} \label{r5}
\end{figure*}

     The system of MHD equations (\ref{eq15})
was integrated numerically
in the region $r_{int}<r<r_{ext}$, $0<\theta
< \pi/2$.
     We used  grids which were equally
spaced in $\theta$.
     The steps in the radial direction were chosen
so that the poloidal plane cells were
curvilinear rectangles with approximately
equal sides.
       A typical region was
$0.5<r<50$, $0<\theta <\pi/2$ with the grid $85\times31$ cells.
Figure \ref{r5} shows the grid in the entire simulation region
(right panel) and in the inner part of the region (left panel).

\begin{figure*}[b]
\epsscale{1.5} \plotone{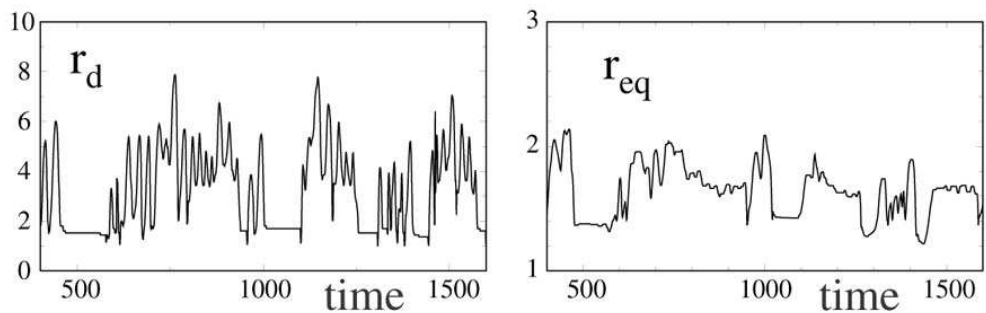} \caption{The left-hand panel shows
the  variations of the inner radius of the disk.
     The right-hand panel shows
the variations of the radius where a given
magnetic field line
    crosses the  equatorial plane.}
\label{r6}
\end{figure*}

In the
{\it hydrodynamic}
        block, the ideal MHD equations are integrated using
an explicit Godunov-type numerical scheme.
       For calculation of fluxes between the
cells we used approximate solution of the Riemann problem
analogous to one described by Brio \& Wu (1988).
      To guarantee the absence of
the magnetic charge, we calculated at each time-step the
$\phi$-component of the vector-potential $A_ {\phi}$.
    This was then used to
obtain the poloidal components of the magnetic field ($B_r$,
$B_\theta$)  in divergence-free form (Toth 2000).

In the block where the
{\it diffusion}
 of the poloidal magnetic field is calculated, we
numerically integrated the equation (\ref{eq20}) for the
$\phi$-component of vector-potential.
      During this calculation we frozen the
values of  $A_\phi$ on the inner and outer boundaries of the
simulation region.
     In the equatorial plane we have conditions of
symmetry, $\partial A_\phi /\partial
\theta=0$.
     On the symmetry axis we have $A_\phi=0$.

     Equation (\ref{eq20}) was approximated
with an implicit difference scheme.
       The approximation was chosen so that the
operator on the implicit time-layer was symmetric and positive.

     For solving the system of equations
on the implicit time-layer, we used ICCG method (Incomplete
Cholesky Conjugate Gradient method).
     Because the  size of the grid cells and
values of the coefficient of the magnetic diffusivity vary
strongly in space, the elements of the matrix of the system also
vary strongly.
    To remove this undesirable property,
we changed the matrix so that it has
diagonal elements equal to unity.

     In the block where the
diffusion of the azimuthal component of the magnetic field was
calculated, we numerically integrated equation (\ref{eq21}).
      At the inner and outer boundaries
the $B_\phi$ was frozen at this computational block.
     Along the rotation axis
and on the equatorial plane $B_\phi=0$.

Equation (\ref{eq21}) was approximated
by a numerical scheme with
symmetric and positive operator on the implicit time layer.
      The corresponding system of linear equation was solved by the
ICCG-method with pre-conditioning.

In the block of our code where
the
{\it viscous}
stress is calculated,  we numerically integrated the equation
(\ref{eq16}) for the angular velocity of matter $\omega = v_\phi / r
\sin \theta$.
      At the inner boundary of the simulation
region we take $\omega=\Omega_*$, the angular rotation
velocity of the star.
     At the outer boundary, $\omega$ is taken to be fixed
and equal to the corresponding Keplerian value.
       On the axis and in the
equatorial plane we have the condition of zero stress for the
$\theta \phi$ -component of the viscous stress tensor.

Equation (\ref{eq16}) was approximated
by a numerical scheme with
the symmetric and positive operator
on the implicit time-layer.
     The corresponding system of  linear equations was solved by the
ICCG-method with pre-conditioning.

\section{Results}

We performed simulations of  accretion to a star in
the ``propeller" regime for a wide range of parameters, $\mu$,
$\Omega_*$, $\alpha_{\rm v}$ and $\alpha_{\rm d}$.
      In the following we discuss
results for the reference case $\mu=10$, $\Omega_*=1$, $\alpha_{\rm
v}=0.2$, and $\alpha_{\rm d}=0.2$.

\begin{figure*}[t]
\epsscale{1.5}
   \plotone{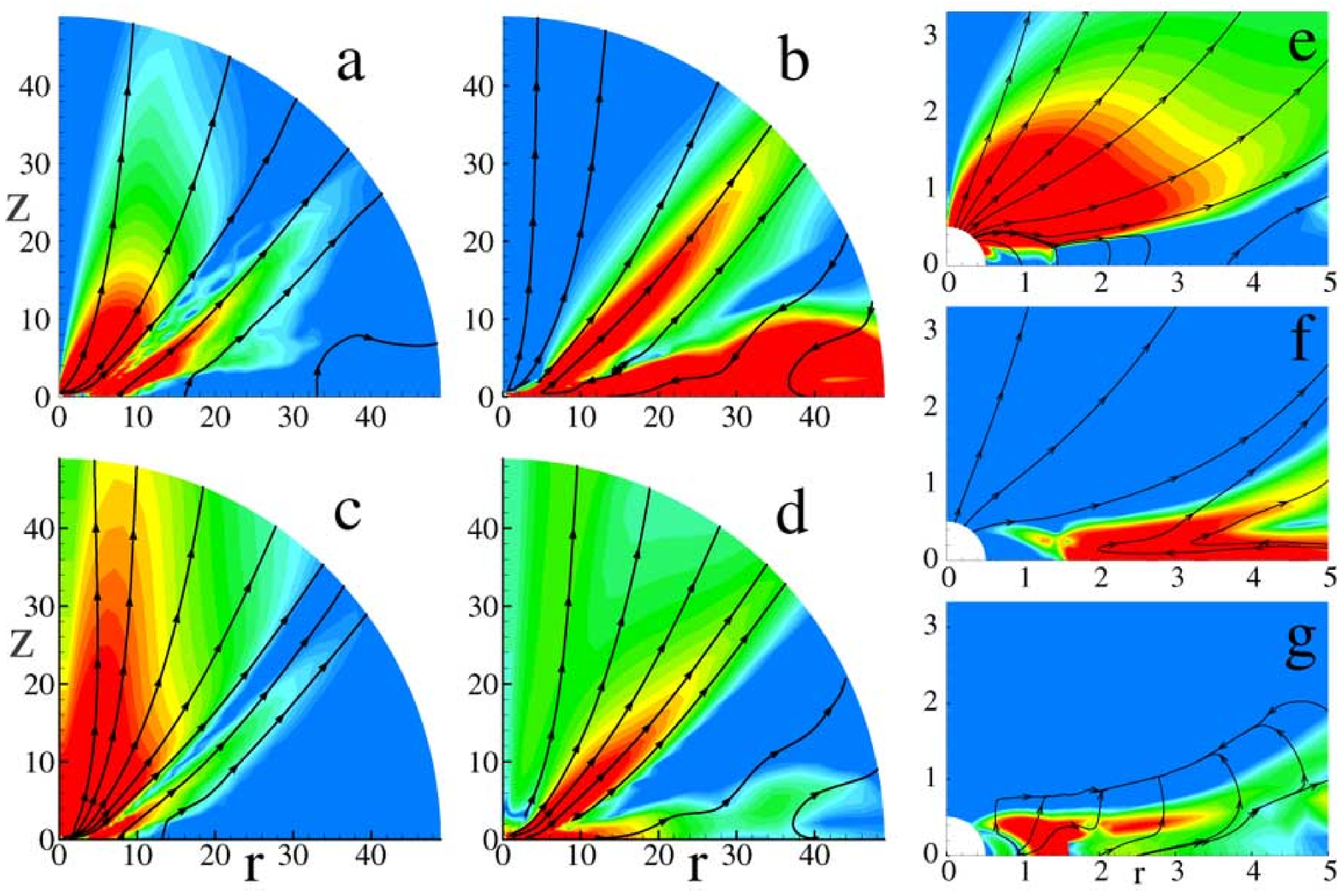}
\caption{The top panels show the angular
momentum flux carried by the
field (a) and  by the matter (b).
     The bottom panels show the energy fluxes
carried by the field (c) and by the matter (d).
     The right-hand panels show the angular
momentum fluxes close to the star carried by the field (e),
by the matter (f),
and by the viscous stress (g).
The background color indicates the
value of the flux while the streamlines
indicate the direction of the flux.}
\label{r7}
\end{figure*}

\begin{figure*}[t]
\epsscale{1.} \plotone{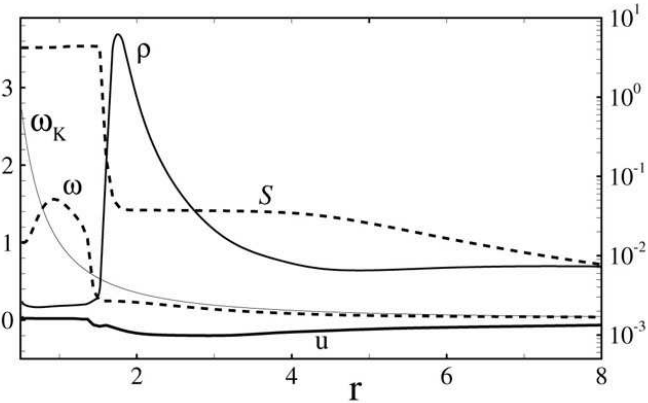} \caption{Radial distribution of
different variables in the equatorial plane, the density $\rho$, the
specific entropy $S$, the radial velocity $v_r$, the angular
rotation rate $\omega=v_\phi/r$, and the Keplerian angular velocity
$\omega_K$.
   The scale for the entropy is on the right-hand side.}
\label{r8}
\end{figure*}

\begin{figure*}[t]
\epsscale{1.3} \plotone{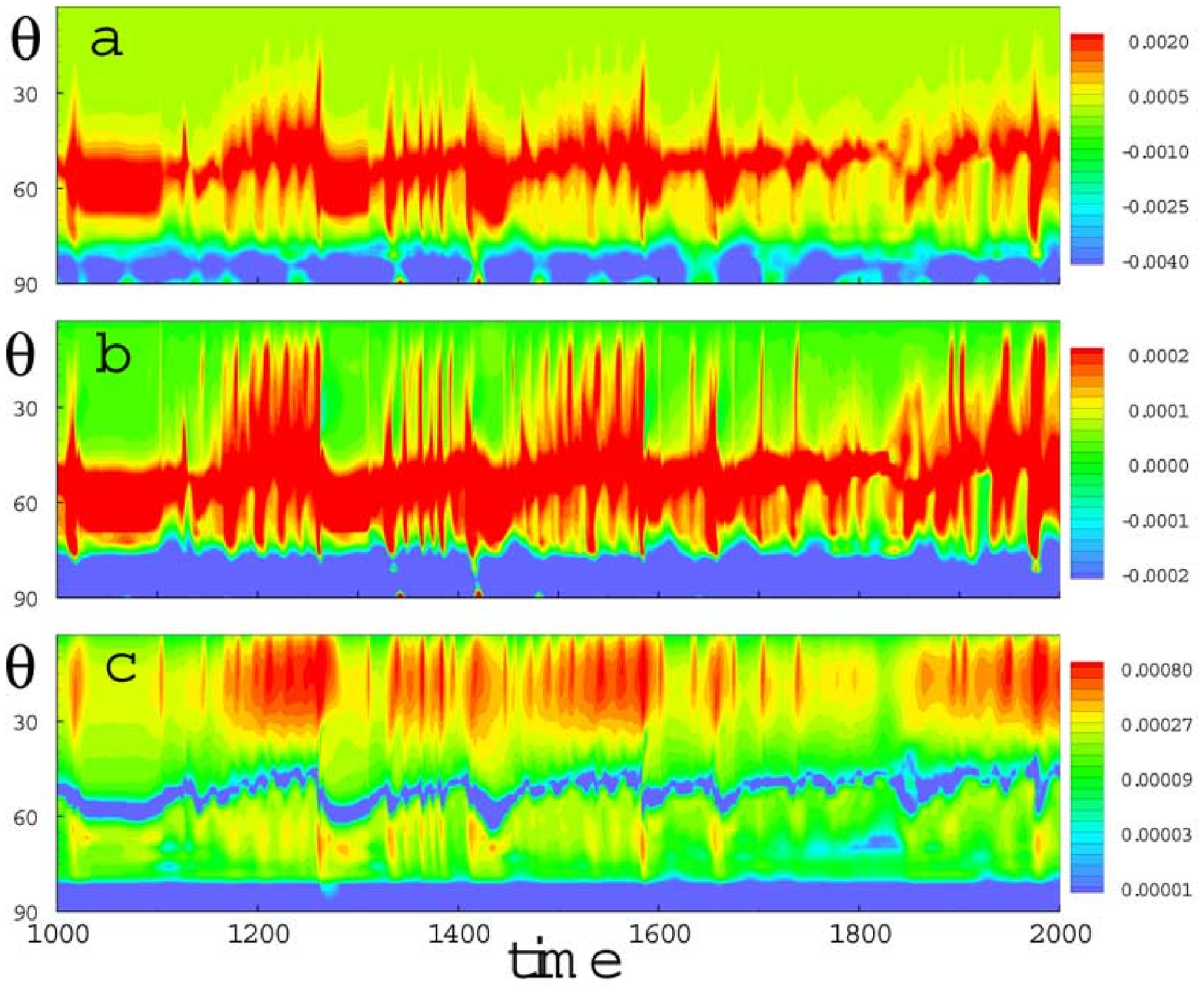} \caption{Angular ($\theta$)
distributions  of different fluxes on the surface of a sphere of
radius $r_0=20$ as a function of time:
    (a) shows the matter flux; and
(b) and (c) show the radial angular momentum flux
carried by the matter and by the magnetic field,
respectively.}
\label{r9}
\end{figure*}

The reference case was calculated
out to a time corresponding to
$2500$  rotation periods of the star.
     We observed that the process of interaction between the
magnetosphere and the disk  is accompanied with
(1) oscillations of the disk, (2) quasi-periodic opening and closing
of the magnetic field, (3) quasi-periodic outbursts to winds
and jets.
     This evolution  arises from the
diffusive mixing of the disk
matter with the magnetic field of the star,
the opening of the magnetic
field lines linking the star
and disk, unloading of the disk matter to the star and to
the outflows, and the outflow of
angular momentum from the star and from the disk.

      The left-hand panel of Figure 7  shows the
variation of the inner radius of the
disk which was calculated at the density level, $\rho_0=0.3$.
Note that near the disk-corona boundary,
$\rho (r , \theta,t )=\rho_0$,
the density gradient is very
large so that the choice of $\rho_0$
is not important.
     During an initial time
$t \sim 100$, part of the magnetic flux outgoing   the star opens
and subsequently varies about an average value.
    To determine this magnetic flux
we calculated the dependence of the maximum value of the magnetic
flux function $\Psi (r,\theta,t )=r \sin \theta A_\phi $,
$\Psi_0={\rm max}_{\theta} \Psi (r_0,\theta,t )$ on time at the
sphere of radius $r_0=30$.
    The magnetic field lines along which
$\Psi > \Psi_0$, are evidently closing inside the sphere of
radius $r_0$.
    The value $\Psi_0$ determines the location of the last
closed field line at this distance, in particular the radius at
which this field line crosses equatorial plane
$r_{eq}=\mu /\Psi_0(t)$. That is, $r_{eq}$ shows the
radius of the closed magnetosphere.
      The right-hand panel of Figure 7  shows
that this radius $r_{eq}(t)$ varies with time in a fashion
similar to $r_d(t)$; however, the amplitude is much smaller.

        An important aspect of the interaction between a
rapidly rotating magnetized star
and an accretion disk is the transport of angular momentum.
     The equation for the conservation of
angular momentum can be obtained from (\ref{eq16}) by multiplying it
by
$r\sin \theta$,
$$
    \frac{\partial( \rho v_\phi r \sin \theta)}{\partial t} +
\frac{1}{r^2} \frac{\partial}{\partial r}
\left\{ r^2 \left[ r \sin \theta
\left( T_{r \phi} - \nu_t \rho r \sin \theta \frac{\partial
\omega}{\partial r} \right) \right]\right\} +
$$
$$
+ \frac{1}{r \sin\theta}
\frac{\partial}{\partial \theta}\left\{ \sin \theta \left[ r \sin
\theta \left( T_{\theta \phi}-\nu_t \rho \sin \theta \frac{\partial
\omega}{\partial \theta} \right) \right]\right\} =0~,
$$
where
$T_{r \phi} = \rho v_r v_\phi - B_r B_\phi /4\pi$ and
$T_{\theta \phi}=\rho v_\theta v_\phi -
B_\theta B_\phi /4\pi$ are the components of ``non-viscous"
part of the stress tensor.

      The angular momentum conservation equation evidently
has the form of a continuity equation,
$$
\frac{\partial (\rho \ell)}{\partial t} +
{\bf \nabla}\cdot {\bf L} =0~,
$$
where
$\ell=v_\phi r \sin \theta$
is the specific angular momentum and the angular momentum flux
density is
$$
{\bf L}= r \sin
\theta \left( \rho v_\phi {\bf v}_p- \frac{B_\phi
{\bf B}_p}{4 \pi} - \nu_t \rho r \sin \theta \mathbf{\nabla}
\omega \right)~.
$$
The first term on the right-hand-side ${\bf L}$
gives  the transport of angular momentum by the matter;
the second term is magnetic field
contribution; and the third term is the transport due to
viscous stress.

Figure 8 (a,b,e-g) shows the angular momentum fluxes at
   $t=2300$.
      The streamlines show the direction of
the fluxes, while the background shows the magnitude
of the flux.
      Figures 8 (a, e) show the distribution
of the angular momentum fluxes
carried by the magnetic field at large and small scales.

 Figures 8(b, f)
show the corresponding fluxes carried by the matter.
      From
Figure 8e
it is clear that the angular momentum outflow from the star
is due mainly to the twisting of the magnetic field.
      Note that a  larger part of
angular momentum is carried along the open field lines. Other part
of angular momentum which is along the closed field lines is
transported from the star to the disk.
       Angular momentum carried by
matter is largest in the disk and in the conical outflow in the
vicinity of the neutral layer (Fig. 8b).
      Here magnetic pressure is
not very high, while pressure and density of matter play larger
role. Bottom panels show energy fluxes carried by the magnetic field
(Fig. 8c) and by matter (Fig. 8d). One can see from Figure 8c that
significant energy is carried from the star by a Poynting flux.

Figure 8g shows the angular momentum
   flux carried by  the viscous
stress.
      This stress is large at the boundary between the
disk and magnetosphere in the vicinity of the equatorial plane. This
is the place of significant gradients of the angular velocity of
matter, as one can see from the Figure 9 (shown at the same time
$t=2300$).
       This figure also shows the  distribution of
density, entropy (logarithm scale at the right), and radial
velocity $v_r$ in the equatorial plane.
      It is evident that the disk has a sharp boundary
both in density and in entropy.

    Next, we evaluate the directions where  most
of the matter and angular
momentum outflows.
      For this we calculated the
distribution of the fluxes  of matter and angular momentum as a
function of $\theta$ (the colatitude) as a function of time at a
given distance $r_0=20$.
     Figure 10 shows a plot of these distributions as
a function of time.
      One can see that the matter flux to the wind
is mainly in the range
$40^\circ <\theta <70^\circ$.
     Thus, most
of the matter flows into a wide-angle hollow-cone with a large
inclination angle to the $z-$axis (Figure 10a).
    There is also a flux of
matter  towards the star
which is associated with
the equatorial region of
the disk.
       The angular momentum
flux carried by matter has distribution  similar to the distribution
of  the matter flux (see Figure 10b).
      The angular momentum flux carried by
the magnetic field (Figure 10c) is concentrated in two stripes:
$0^\circ <\theta <30^\circ$
(near the axis), and
$60^\circ <\theta<80^\circ$
(closer to the disk).
     The larger flux is associated
with the angular momentum outflow  in
collimated jet, along the field
lines which start at the surface of the star.
The other stripe, at larger angles
$\theta$ is connected with the disk wind.

    The wide-angle, hollow-cone disk wind
is  similar to the stationary magnetocentrifugally
driven winds predicted by Blandford and Payne (1982)
and Lovelace, Berk, and Contopoulos (1991) and
first obtained and analyzed  in MHD simulations by
Ustyugova et al. (1999).  For such winds the
angle of the poloidal field to the $z-$axis
at the surface of the disk $\Theta =
\tan^{-1}(|B_r/B_z|)$
is predicted to be  $\gtrsim 30^\circ$.
    On the other hand  a magnetically
dominated (or Poynting) outflow from a disk or
star can have the angle $\Theta$ less than
$30^\circ$ (Ustyugova et al. 2000; Lovelace et al. 2002).
    The narrow-angle, hollow-cone jet along
the open field lines of the star has
much smaller values of  $\Theta$ and it is
magnetically dominated.

\begin{figure*}[t]
\epsscale{1.5} \plotone{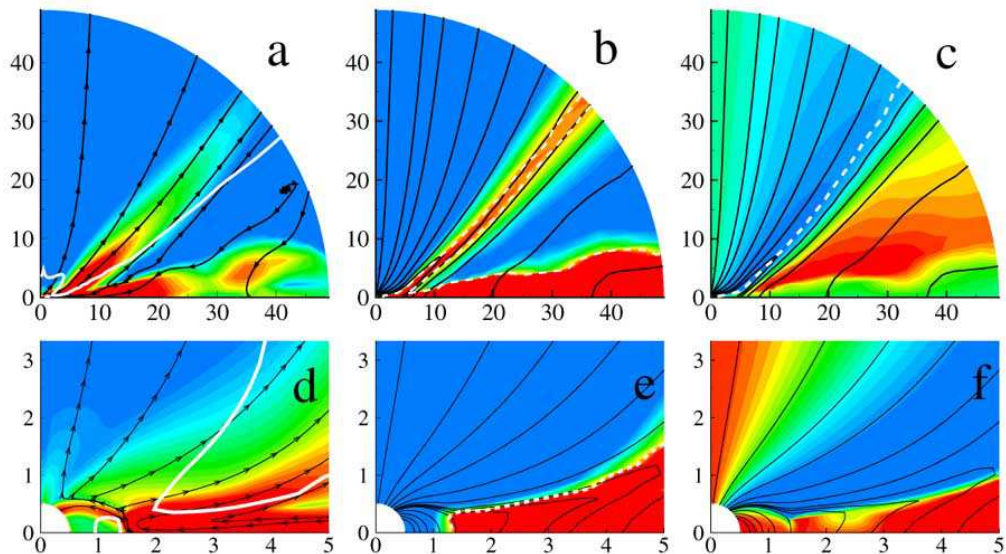} \caption{In panel (a) the
background shows the matter flux and the lines are the streamlines
of the matter flow. The white line is the $v=v_{esc}$ line. In panel
(b) the background shows the plasma parameter $\beta=8\pi p/{B^2}$.
    The   streamlines are the magnetic field lines,
and the white dashed line
      shows the $\beta=1$ line.
In panel   (c) the background shows  $r
\sin\theta B_\phi$, which
for steady conditions
is the total poloidal current through a disk $\leq r\sin \theta$
center on the $z-$axis.
     The lines are the magnetic field lines.
      The white
dashed line shows the neutral line where $B_p=0$.
   The bottom panels (d - f) show the same as (a - c)
but in the vicinity of the
star.  }
\label{r11}
\end{figure*}

Figure 11 shows the distribution of different physical quantities in
the computational region (top panels) and on a smaller scale (bottom
panels). Figures 11 (a,~d)
    show distribution of the matter
flux $\rho v_p$, and streamlines of matter
flow at time $t=2300$.
      One can see that matter flows with the
disk, then turns near the magnetosphere
of the star and flows outward in
a disk wind.
    Some matter goes
around the magnetosphere and flows
to the jet along the field lines
starting at the stellar surface.
     The white line shows the escape velocity
$v_{esc}=\sqrt{2GM /r}$. Above this line $v > v_{esc}$. The color
background of
 Figures 11 (b, e)
shows the plasma $\beta\equiv 8 \pi p/{B^2}$ parameter.
      The dashed white
line corresponds to $\beta=1$.
Figures 11 (c, f)
show isocontours of the  $r \sin\theta B_\phi$ which
is the poloidal current flow through a
horizontal disk $\leq r\sin\theta$ centered on the $z-$axis.

\begin{figure*}[t]
\epsscale{1.}
\plotone{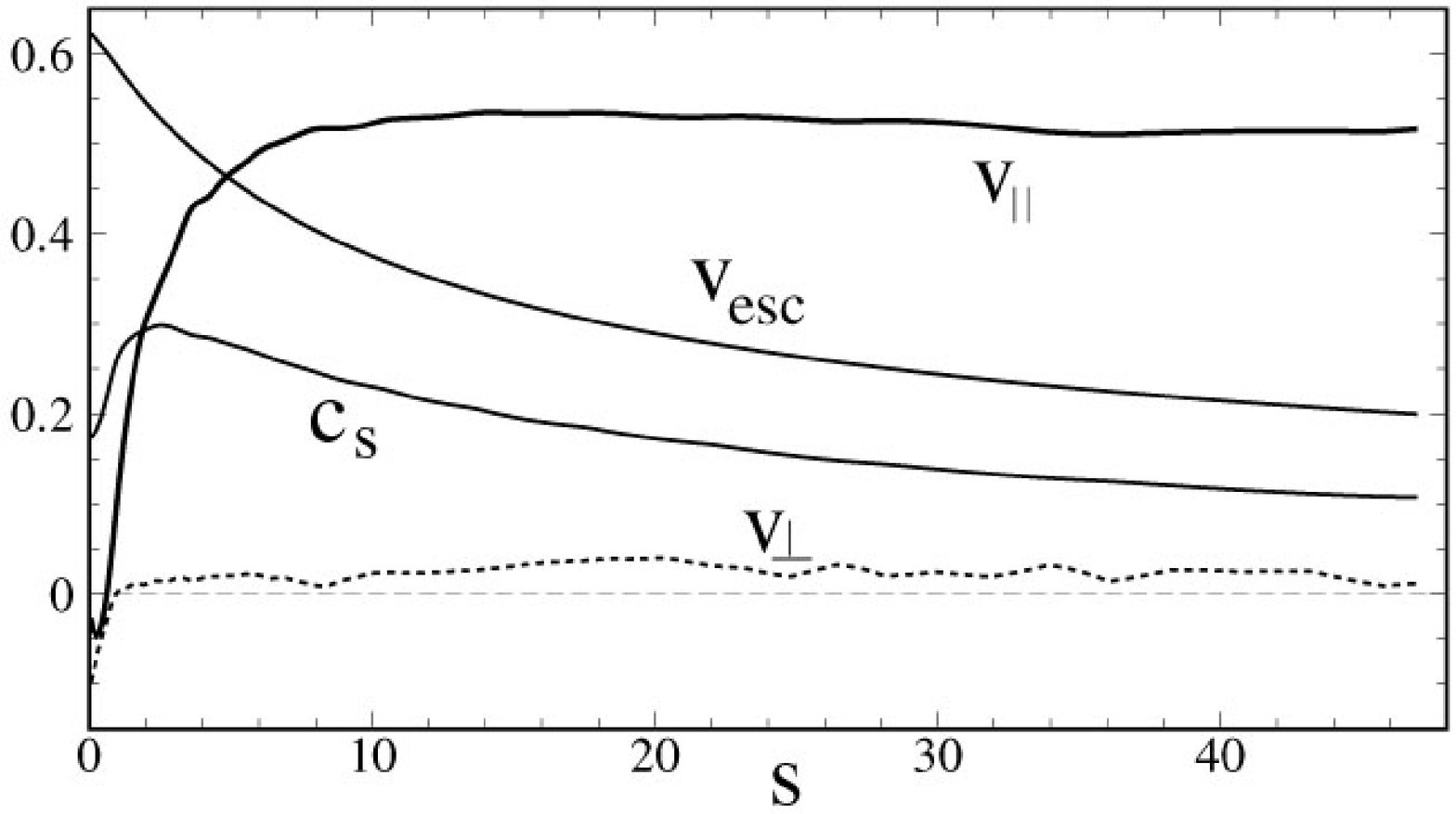}
\caption{Velocities along a field line which starts
from the disk near the neutral
line of the magnetic field (where the matter flux to the
wind is the largest).
$v_{\parallel}$ is the velocity component
parallel to the field line and $v_\perp$ is the component
perpendicular to the field line. $v_{esc}$ is escape velocity
and $c_s$ is a sound speed.}
\label{r12}
\end{figure*}

Figure 12 shows plots of different
variables as a function of
distance along a magnetic  field
line which starts from the disk
in the vicinity of the neutral field line
where the matter flux to the wind is
the largest.
    The solid line shows the projection
of the poloidal velocity onto
this field line, the dashed line
shows the projection of the velocity
component normal to the magnetic field line.
    One can see that the
poloidal velocity is nearly
parallel with the poloidal field,
which is in accord with the theory
of stationary axisymmetric flows
of ideally conducting plasma
(see, e.g., Ustyugova et al. 1999).
    The sound speed $c_s$ and escape
velocity $v_{esc}$ are also shown.
   One can see that the flow
becomes supersonic and that the
flow velocity exceeds the escape
velocity $v_{esc}$.
   Earlier, Ustyugova et al. (1999)
obtained dependences similar to those of Figure 12 for
the case of an ordered magnetic field threading
a disk around a non-magnetized star.

   Interaction of the magnetized star
with the accretion disk can lead to
the outflow and escape
of a significant fraction of the matter incoming
in the disk.
    We find in general that there is a {\it jet} and a
{\it disk wind} giving outflows of energy, angular momentum, and
matter from the system.
     The {\it jet} is identified as the flow
in the region within
the neutral line of the poloidal magnetic field.
The magnetic flux in the jet is the flux emanating
from the star.
     The {\it disk wind} is the flow outside of the
neutral line of the poloidal magnetic field.
     For both outflows we take into account
only the matter which has a velocity higher than the escape velocity
$v_{esc}$.

    Thus, the simulation region includes three flow regions,
the disk, the jet, and the wind.
    These regions do not constitute
the entire simulation volume because part of
it is filled with relatively rarefied
matter of magnetosphere and corona.
     Accreting matter which is not
accelerated to  escape speed goes into the magnetosphere
or the corona.
    These regions do not have a  significant
role in the matter and angular momentum flow.
      The star has an important role
in absorbing mass and losing angular momentum.
    The star influences the plasma dynamics  through
the inner boundary of the simulation region. However the influence
of the plasma on the star is
noticeable
on only an extremely long time scale. Figure 13 shows the
interactions between the different elements of the system.
Evidently, matter which comes through an accretion disk flows in
part to the disk wind, in part to the jet, and in part accretes to
the star. There is also angular momentum transport between different
elements. The star represents a large reservoir of angular momentum
which is transported to the disk through the closed field lines and
to the jet through the open field lines. The positive direction of
the fluxes is shown by arrows.

\begin{figure*}[t]
\epsscale{1.2} \plotone{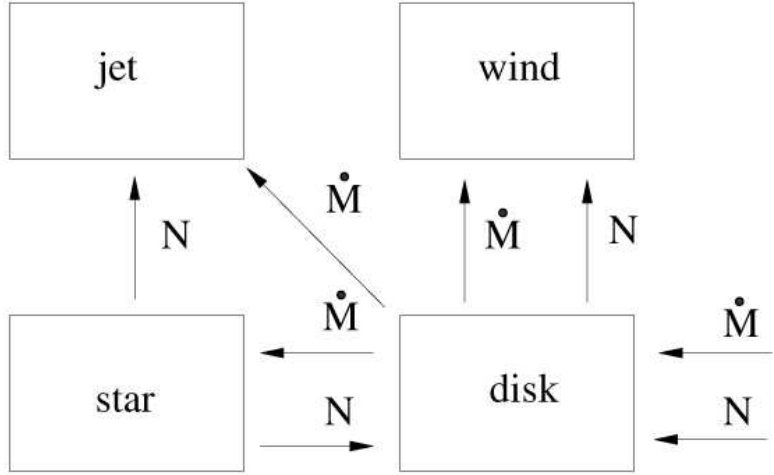}
      \caption{Scheme of interaction between
different elements of the system.}
      \label{r13}
\end{figure*}

\begin{figure*}[t]
\epsscale{1.} \plotone{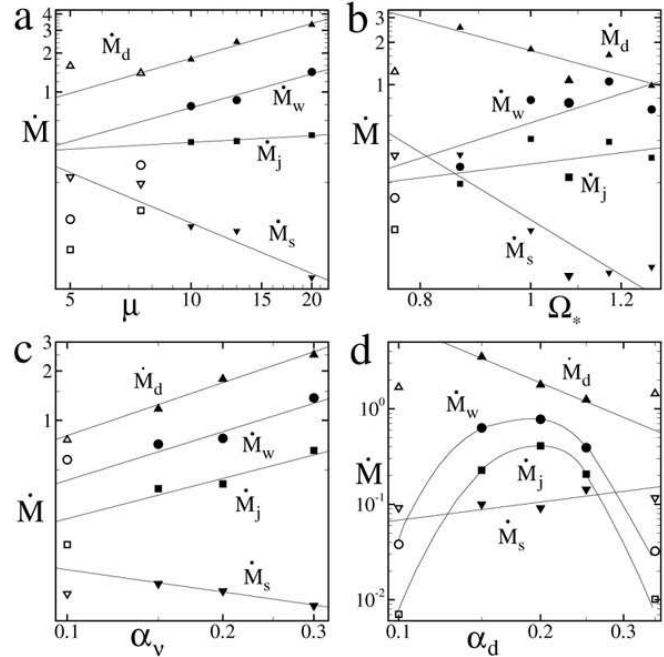}
  \caption{Variation of the matter
fluxes as a function of different parameters.}
\label{massa}
\end{figure*}

In all cases the fluxes of matter and angular momentum
oscillate strongly in time, but always vary around an average value.
    See for example Figure 10.
     For this reason we calculated the time-averaged
fluxes and investigated  their dependences  on the
parameters of the model.
     We took as a base the mentioned reference case with
dimensionless parameters $\mu = 10$, $\Omega_*=1$, $\alpha_{\rm
v}=0.2$, and $\alpha_{\rm d}=0.2$.
    A series of
simulations was then done to obtain
the dependences  of the time-averaged fluxes on
different parameters.  Only one parameter
was varied in each series.
   Figures 14-16 show the derived dependences.

Figure 14 (a-d)
show the dependences
of the matter fluxes on our main
parameters.
   Here, $\dot M_d$ is the mass accretion rate in the
disk at the radius
$r_0=30,$
 $\dot{ M}_s$ is the accretion rate to the star, $\dot M_w$ is the
matter flux to the wind, $\dot M_j$ is the matter flux to the jet.
    As we mentioned, we distinguish between
``strong"
(with outflows) and ``weak"
(no outflows)
propellers.
    Results of modeling
of the ``strong" propellers are marked
with the filled symbols, while
the ``weak" propeller results
are indicated by open symbols.
    Different
symbols show different dependences:
$\dot{M}_j$ (squares),
$\dot{M}_w$ (circles),
$\dot{M}_d$ (triangles), and $\dot{M}_s$
(gradient signs).
    The solid lines show approximations of these
dependences.
   Most of dependences could be approximated by a
simple power law and have a clear sense.
For example, the larger the magnetic moment
$\mu$ and the larger  the angular velocity
of the star $\Omega_*$, the larger
matter fluxes to the wind and jet
and the smaller the accretion rate
to the star.
   At smaller $\mu$ and $\Omega_*$,
ejection of matter to the wind and jet becomes less efficient and we
enter the ``weak" propeller regime with very weak or no outflows
(see Fig. 14 a,b).
   Note, that all ``weak" propellers
are on the left-hand side of the plots.

    It is interesting to look at dependences on viscosity,
$\alpha_{\rm v}$ (see Fig. 14c).    We see that matter fluxes to the
wind/jet strongly increase with increase of viscosity, while the
accretion rate to the star decreases.
   We conclude that the viscous stress
contributes to launching the outflows.
     At sufficiently small
viscosity, $\alpha_{\rm v} \lesssim 0.1$, outflows are absent.
    The role of the viscosity is
twofold. From one side, at larger
viscosity, the ``friction" between
the inner regions of the disk and magnetosphere is larger.
   From other side, the radial
velocity in the disk is proportional to $\alpha_{\rm v}$, so that
viscosity increases the matter flux per unit area and the inner
region of the disk come closer to the inner regions of the fast
rotating magnetosphere. Both factors lead to enhanced outflows, but
it is difficult to separate the two factors.
In RUKL05
test simulations were done for a case with $\alpha_{\rm v}=0.1$, and
with the density in the disk doubled.
    This of course
increased
the accretion rate but it also enhanced the outflows.

    Figure 14d shows the dependence
on the  magnetic diffusivity.
    The
dependence of the matter fluxes to
the wind/jet on $\alpha_{\rm
d}$ is more complicated than other dependences.
    Namely, for
$\alpha_{\rm d} \lesssim 0.2$,
the fluxes $\dot M_w$ and $\dot M_j$
decrease with $\alpha_{\rm d}$,
because the mixing of the disk
matter with the magnetic field
of the magnetosphere become less and
less efficient, and correspondingly
the angular momentum transport from the star to the disk matter
decreases.
   On the other hand, for
$\alpha_{\rm d}\gtrsim 0.2$, the
diffusivity become too high,
and the inner regions of the disk and
magnetosphere are not coupled
sufficiently to transport angular
momentum.
    This explains the parabolic approximations of these
dependences.
    We took into account all dependences shown in
Figures 14 (a-d), and approximated
them with analytic functions.
    We approximated the dependences
on $\alpha_{\rm d}$ with  power laws
to the left and to the right of the dividing value
$\alpha_{\rm d} \approx 0.2$.
   In Figures 14 (c-d), we show the
dependences around $\alpha_{\rm v}\sim 0.2$ and $\alpha_{\rm
d}\sim 0.2$.
    We have made additional runs  not shown in the
plot at much smaller $\alpha_{\rm v}$ and $\alpha_{\rm d}$, down to
$0.01$.
     Also we  have made runs for much larger values, up to $0.6$.
   We show below dependences which incorporated all these runs,
separately, for $\alpha_{\rm d}>0.2$ and $\alpha_{\rm d} < 0.2$.
   The matter fluxes to the wind $\dot M_w$, to the jet $\dot M_j$,
through the disk $\dot M_d$, and to the  star $\dot
M_s$ were approximated as:
$$ \dot M_w= 0.8 {\Omega_*}^{2.6}
\left(\frac{\mu}{10}\right)^{0.9} \left(\frac{\alpha_{\rm
v}}{0.2}\right)
    \left \{
      \begin{array}{ll}
\displaystyle{\left(\frac{\alpha_{\rm d}}{0.2}\right)^{-4.4}} &
        \alpha_{\rm d} > 0.2 \\[0.3cm]
\displaystyle{\left(\frac{\alpha_{\rm d}}{0.2}\right)^{2}} &
\alpha_{\rm d} < 0.2
      \end{array}
\right.,
$$
$$ \dot M_j= 0.4 {\Omega_*}
\left(\frac{\mu}{10}\right)^{0.2} \left(\frac{\alpha_{\rm
v}}{0.2}\right)^{0.8}
    \left \{
      \begin{array}{ll}
\displaystyle{ \left(\frac{ \alpha_{\rm d}}{0.2}\right)^{-3.6} }&
\alpha_{\rm d} > 0.2 \\[0.3cm]
\displaystyle{\left(\frac{\alpha_{\rm d}}{0.2}\right)^{1.7} }&
\alpha_{\rm d} < 0.2
      \end{array}
\right.,
$$
$$ \dot M_s= 0.1 {\Omega_*}^{-5}
\left(\frac{\mu}{10}\right)^{-1.3}
    \left(\frac{\alpha_{\rm v}}{0.2}\right)^{-0.5}
\left(\frac{\alpha_{\rm d}}{0.2}\right)^{0.6},
$$
$$ \dot M_d= 1.8 {\Omega_*}^{-2.2}
\left(\frac{\mu}{10}\right)^{0.9}
    \left(\frac{\alpha_{\rm v}}{0.2}\right)
\left(\frac{\alpha_{\rm d}}{0.2}\right)^{-2.1}.
$$

      The matter flux
to the disk wind increases with $\mu$
almost linearly and strongly increases with $\Omega_*$, while the
accretion rate to the star $\dot M_s$ strongly decreases.
We note that dependences on $\mu$ and $\Omega_*$
have a threshold character.
    Namely, the matter fluxes to the wind and to the jet
strongly decrease for $\mu \lesssim
7$ and $\Omega_*\lesssim 0.8$,
where the system enters the ``weak"
propeller regime (see open symbols in Figures 14 a,b).
   One can see
from the Figure 14b that the dependence of $\dot{M}_w$ and
$\dot{M}_j$ on the angular velocity of the star $\Omega_*$ also has
a threshold behavior for $\Omega_* \lesssim 0.8$.
   These threshold dependences reflect the
qualitative difference between
``strong" propellers (with outflows) and ``weak" propellers (with
no outflows).

As a measure of the efficiency of the propeller
interaction  with the  disk, we
take the ratio of the  outflow rate to the  disk wind and jet
to the accretion rate through the disk,
  ${\cal S} =\dot{M}_{jw} /\dot{M}_d$.
    For the larger values of $\cal S$, less
matter reaches the surface of the star. Correspondingly, a larger
part of the  disk matter acquires additional angular momentum and
flows to the jet and/or wind.
   Figures 15 a-d show the dependences of the value $\cal S$ on
the parameters of the model.
   The efficiency of the propeller depends weakly on the
magnetic moment of the star.
    The larger the angular
velocity of the star and the larger the $\alpha$-coefficient of
viscosity, the higher is the efficiency of the propeller.
    The dependence of the efficiency on the
$\alpha$-coefficient of  magnetic diffusivity has
a maximum near $\alpha_{\rm d} \sim 0.2$.
    We found the following
dependences,
$$
\frac{\dot M_{jw}}{\dot M_d} = 0.7 {\Omega_*}^{5.6}
\left(\frac{\mu}{10}\right)^{-0.2}
\left(\frac{\alpha_{\rm v}}{0.2}\right)^{1.3}
    \left \{
      \begin{array}{ll}
\displaystyle{ \left(\frac{ \alpha_{\rm d}}{0.2}\right)^{-4.7} }&
\alpha_{\rm d} > 0.2 \\[0.3cm]
\displaystyle{ \left(\frac{\alpha_{\rm d}}{0.2}\right)^{1.7} }&
\alpha_{\rm d} < 0.2~.
      \end{array}
\right.
$$
Another measure of the efficiency of the propeller is
the ratio of the total matter flux going to the wind/jet to the
matter flux to the star,
$$
\frac{\dot{M}_{jw}}{\dot{M}_s} = 13 {\Omega_*}^{10}
\left(\frac{\mu}{10}\right)^{2}
\left(\frac{\alpha_{\rm v}}{0.2}\right)^{2.5}
    \left \{
      \begin{array}{ll}
\displaystyle{\left(\frac{\alpha_{\rm d}}{0.2}\right)^{-3.9} }&
\alpha_{\rm d} > 0.2 \\[0.3cm]
\displaystyle{\left(\frac{\alpha_{\rm d}}{0.2}\right)^{2.1} }&
\alpha_{\rm d} < 0.2~.
   \end{array}
\right.
$$
The ratio of outflow to accretion increases strongly  with
$\mu$ and $\Omega_*$.

\begin{figure*}[t]
\epsscale{1.}
  \plotone{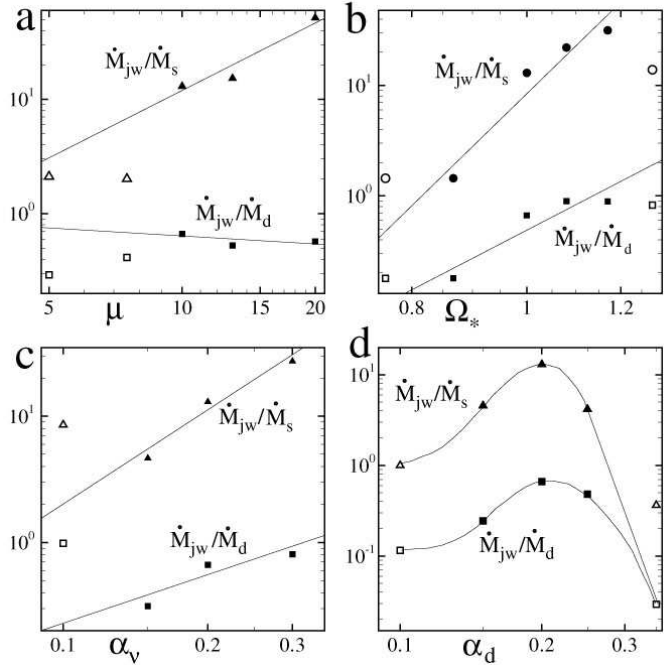}
\caption{Efficiency of the propeller as
a function of different parameters as discussed in the
text.}
\label{eta}
\end{figure*}

\begin{figure*}[t]
\epsscale{1.} \plotone{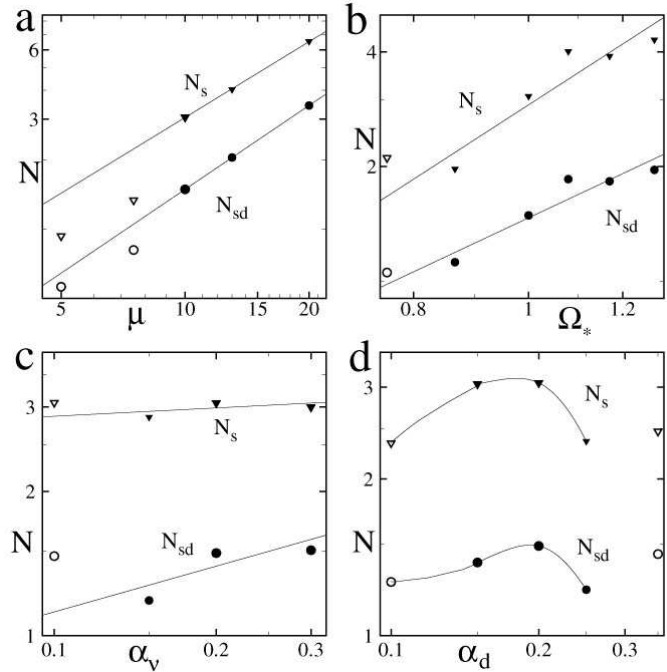} \caption{Variation of the angular
momentum fluxes 
with parameters of the system .} \label{angl}
\end{figure*}

Next, we discuss angular momentum transport
from the star
and its dependence on main parameters of the model.
   The outflow of
angular momentum from the star causes it to spin down.
   Part of the angular momentum is
carried  by the matter,  another
part by the tension of the
magnetic field lines,
and a further part  by the viscous stress.

   The flux of angular momentum
from the star was calculated as
$$
N_s = \int d {\bf S}\cdot r \sin \theta \left[ \rho v_\phi {\bf v}_p
- \frac{B_\phi {\bf B}_p}{4 \pi}
- \nu_t \rho r \sin \theta \nabla \omega \right]~,
$$
where the integration was over the surface of the inner boundary of
the simulation region which is close to the surface of the star and
where $d {\bf S}$ is the surface element directed outward to the
region.
Simulations show that close to the star angular momentum flux is
determined mainly by the magnetic stress.

Part of the flux is associated with the closed field lines. These
are field lines connecting a star and the disk. Other part is
associated with the open field lines, which connect a star with the
low-density corona. We calculated a total flux $N_s$ and also a flux
associated with the closed field lines, $N_{sd}$, that is angular
momentum flux transported from the star to the disk.

Figure 16 shows fluxes ${N}_s$ (gradient signs) and ${N}_{sd}$
(circles). We found the following dependences on different
parameters:
$$
N_{sd} =  1.5 {\Omega_*}^{1.5}
\left(\frac{\mu}{10}\right)^{1.2} \left(\frac{\alpha_{\rm
v}}{0.2}\right)^{0.3}
    \left \{
      \begin{array}{ll}
\displaystyle{ \left(\frac{\alpha_{\rm d}}{0.2}\right)^{-0.2} }
& \alpha_{\rm d} > 0.2 \\[0.3cm]
\displaystyle{\left(\frac{\alpha_{\rm d}}{0.2}\right)^{0.14} }&
\alpha_{\rm d} < 0.2~,
      \end{array}
\right.
$$
$$ N_s =  3.1 {\Omega_*}^{2}
\left(\frac{\mu}{10}\right)^{1.1}  \left(\frac{\alpha_{\rm
v}}{0.2}\right)^{0.1}
    \left \{
      \begin{array}{ll}
\displaystyle{\left(\frac{ \alpha_{\rm d}}{0.2}\right)^{-0.4}} &
\alpha_{\rm d} > 0.2 \\[0.3cm]
\displaystyle{\left(\frac{\alpha_{\rm d}}{0.2}\right)^{0.5} }&
\alpha_{\rm d} < 0.2~.
      \end{array}
\right.
$$

Figure 16 (a, b) shows that both fluxes increase with
  $\mu$ and $\Omega_*$.
They increase with a similar rate so that the lines in Figures 16
(a,b) are almost parallel.
   Figures 16 (c,d) show that the
fluxes depend only weakly on the $\alpha$ coefficients of viscosity
and diffusivity.
They
 have maximum at $\alpha_{\rm d} \sim 0.2$.
    Note that the angular momentum
flux carried by the open field lines of the jet
 $N_s - N_{sd}$ is similar to
that carried by the closed field lines. That is, a star in the
propeller regime in our range of parameters spins-down
due to both: open and closed field lines.

Now, we can estimate the time-scale of spin-down for CTTSs and
accreting neutron stars for parameters given in \S 3.4. For CTTSs,
the loss of the angular momentum of the star in our main case is
$N_s=3.1 \dot L_0 = 1.06\times 10^{37} {\rm g cm}^2/{s}^2$.
  The star's angular
velocity is $\Omega=2\pi/P \approx 7\times 10^{-5} {\rm s}^{-1}$,
its angular momentum is $J= k M r^2  \Omega =
 2.2\times  10^{51} k ~{\rm g cm}^2/{\rm s}$, where $k<1$.
   Taking $k=0.4$, the spin-down
time-scale is $\tau = J/N_s \approx 2.7\times 10^6$ years. Note,
that this time-scale was calculated for a magnetic field $B_*=10^3
{\rm G}$ and a relatively low accretion rate, $\dot M\approx
10^{-8}{\rm M_\odot/yr}$. The time-scale decreases with magnetic
field of the star as $\sim B^{-1.1}$ and will be $\tau \approx
8.1\times 10^5$ years for $B_*=3\times 10^3 {\rm G}$. Time-scale
also decreases with matter flux. Thus, if CTTSs have a strong
magnetic field in the past, then they were at the propeller regime
and were spun-down in $ < 10^6$ years. This time-scale is shorter
than typical life-time of the CTTSs, which is $\sim 10^7-10^8$
years. We conclude that propeller mechanism may be responsible for
fast spinning-down of CTTSs to presently observed slow rotation in
the early stages of their evolution.

    For an accreting neutron star, $N_s=3.1 \dot L_0 = 3.87\times 10^{33}
{\rm g cm}^2/{\rm s}^2$,
 $\Omega=2 \pi /P \approx 5\times 10^3 {\rm s^{-1}}$.
   The star's  angular momentum
 is $J = k  M  r^2  \Omega =
 1.5\times 10^{49} k~ {\rm g cm^2/s}$.
The time-scale of spin-down is $\tau = J/N_s \approx 2.5\times 10^7$
years. This time-scale is shorter than expected life-time of
millisecond pulsars in the accreting stage. However, millisecond
pulsars have a different history of evolution compared to CTTSs, and
they expected to be spined up during accreting stage
(Bisnovatyi-Kogan \& Komberg 1974). However, the pulsar mechanism
may be responsible for fast spinning-down of the pulsar during
periods of decreased accretion rate.

\section{Conclusions}

    We performed a large number
of axisymmetric MHD simulations and
did extensive analysis
of the interaction of
a rapidly rotating magnetized
star with an accretion disk
in the propeller regime.
The results can be divided to two classes, ``strong"
propellers and ``weak" propellers.
    The ``strong" propeller is characterized by the intense,
wide-angle, conical
matter outflow in the vicinity of the neutral line of the
poloidal magnetic field and formation of well collimated,
magnetically dominated jets  along the rotational axis.
   In the
``weak" propeller regime there are no appreciable matter outflows
from the disk and the collimated  flow along the axis is much
weaker.

In this work the four main
parameters of the model were
varied:
the
magnetic moment of the star, its angular velocity, and $\alpha-$
coefficients of the kinematic viscosity and magnetic diffusivity.
   We calculated time-averaged total
fluxes of matter, energy, and
angular momentum and investigated
their dependences on the main parameters
of the model.
    The average characteristics were calculated  during
$1000 - 2500$  periods of rotation of the inner disk.
    The derived dependences were approximated in most
cases by power law functions. But some dependences showed a
threshold character with the flux becoming negligible for a range of
the parameters.

In all cases the interaction of the
magnetized star with the accretion
disk showed strongly non-stationary,
quasi-periodic character.
   For example, in
the reference case described in the paper,  the typical ``period"
of oscillations is several tens of our time-units.
  All fluxes of
matter, energy, and angular momentum to the wind, jet, and star
oscillate strongly with time.

Our simulation results are applied to classical T Tauri stars and to
millisecond pulsars. We conclude that propeller mechanism may be
responsible for fast spinning-down of CTTSs to presently observed
slow rotation during the first $\sim 10^6$ years of their evolution.
In the case of accreting, magnetized millisecond pulsars, the
propeller mechanism may be responsible for relatively fast spin-down
during periods of lower accretion.

\acknowledgments This work was supported in part by NASA grants
NAG5-13220, NAG5-13060, NNG05GG77G,  by NSF grants AST-0307817,
AST-0507760 and by the CRDF grant KP2-2555-AL-03. AVK and GVU were
partially supported by
the grant RFBR 06-02-16608.

\end{document}